\RequirePackage{fix-cm}
\documentclass[twocolumn,epjc3]{svjour3}
\smartqed
\RequirePackage{graphicx}

\usepackage{amssymb,graphics}
\usepackage{graphics,epsfig,subfigure}
\usepackage{epstopdf}
\usepackage{graphicx}
\usepackage{dcolumn,color}
\usepackage{bm}
\usepackage{epsfig}
\usepackage{color,framed}
\usepackage{amsmath,amssymb,graphics,epsfig,subfigure}

\journalname{Eur. Phys. J. C}

\begin{document}
\renewcommand{\baselinestretch}{1.3}
\newcommand\beq{\begin{equation}}
\newcommand\eeq{\end{equation}}
\newcommand\beqn{\begin{eqnarray}}
\newcommand\eeqn{\end{eqnarray}}
\newcommand\fc{\frac}
\newcommand\lt{\left}
\newcommand\rt{\right}
\newcommand\pt{\partial}

\title{Gravitational resonances on $f(R)$-brane}

\author{Hao Yu\thanksref{e1,addr1}
        \and Yuan Zhong\thanksref{e2,addr1,addr2}
         \and Bao-Min Gu\thanksref{e3,addr1}
         \and Yu-Xiao Liu\thanksref{e4,addr1,addr3}
        }

\thankstext{e1}{e-mail:yuh13@lzu.edu.cn}
\thankstext{e2}{e-mail:zhongy2009@lzu.edu.cn}
\thankstext{e3}{e-mail:gubm09@lzu.edu.cn}
\thankstext{e4}{e-mail:liuyx@lzu.edu.cn, corresponding author}

\institute{Institute of Theoretical Physics,
            Lanzhou University, Lanzhou 730000,
            People's Republic of China \label{addr1}
           \and
           IFAE, Universitat Aut$\grave{\textrm{o}}$noma de Barcelona, 08193 Bellaterra, Barcelona, Spain \label{addr2}
           \and
           Key Laboratory for Magnetism and Magnetic Materials of the MoE,
                 Lanzhou University, Lanzhou 730000, China \label{addr3}
}

\date{Received: 23 December 2015 / Accepted: 28 March 2016}

\maketitle

\begin{abstract}
{In this paper, we investigate various $f(R)$-brane models and compare their gravitational resonance structures with the corresponding general relativity (GR)-branes. {Starting from some known GR-brane solutions}, we derive thick $f(R)$-brane solutions such that the metric, scalar field, and scalar potential coincide with those of the corresponding GR-branes. {We find that for branes generated by a single or several canonical scalar fields, there is no obvious distinction between the GR-branes and corresponding $f(R)$-branes in terms of gravitational resonance structure.} Then we discuss the branes generated by K-fields. In this case, there could exist huge differences between GR-branes and $f(R)$-branes.}
\end{abstract}

\section{Introduction}

Since general relativity (GR) was established by Einstein in 1915, a wide range of new theories of gravity have been proposed in the past 100 years. The simplest way to modify GR is to enlarge the dimensions of the space-time. The Kaluza-Klein (KK) theory as one of the pioneering five-dimensional space-time theories was proposed to unify electromagnetism with GR in the 1920s by Kaluza and Klein~\cite{Kaluza1921,Klein1926}. In order to generate effective four-dimensional theories, the extra spatial dimensions in the KK types of theories are assumed to be compacted to the Planck scale. {About 60 years later, physicists found that if our world is some kind of four-dimensional domain wall in a higher-dimensional space-time, then the matter fields can be dynamically trapped on the wall.} As a consequence, our world is effectively four-dimensional, even for infinite extra dimensions~\cite{Akama1982,RubakovShaposhnikov1983,RubakovShaposhnikov1983a,Visser1985,Squires1986}. {String theory also predicts the existence of TeV scale extra dimensions~\cite{Antoniadis1990}.}

But what really triggered the modern revolution of extra dimension theories is the work done by Arkani-Hamed, Dimopoulos, and Dvali (ADD)~\cite{Arkani-HamedDimopoulosDvali1998a}, which provides an alternative solution to the gauge hierarchy problem. {The idea of ADD model is that all the matter fields are confined on the four-dimensional brane embedded in a higher-dimensional space-time, while gravity can propagate in the bulk.} In ADD model, the fundamental scale is assumed to be the electroweak scale $M_{\ast}=m_{\textrm{EW}}\sim 1$ TeV, and the large hierarchy between the Planck scale $M_{\textrm{Pl}}$ and the fundamental scale is ascribed to the large volume of the extra dimensions. Due to the compactification of the extra dimensions, the four-dimensional Newton law appears at a distance much larger than the radius of the extra dimensions.

Soon after the ADD model, Randall and Sundrum (RS) found that by using a nonfactorizable warped geometry, one can also solve the hierarchy problem~\cite{RandallSundrum1999}. More interestingly, they have shown that it is possible to generate the four-dimensional Newtonian law at large distances in a warped space-time even when the extra dimension is infinitely large~\cite{RandallSundrum1999a}. Now, these two models are known as the RS-I and the RS-II models, respectively. In the RS-II model, the four-dimensional Newtonian law is generated by a gravitational zero mode localized near the brane. Later, Gregory, Rubakov, and Sibiryakov (the GRS model)~\cite{GregoryRubakovSibiryakov2000b} have shown that four-dimensional gravity can also be reproduced at an intermediate scale even without a localizable gravitational zero mode. In fact, four-dimensional Newtonian gravity is generated by a quasilocalized graviton (or gravitational resonance) in the GRS model~\cite{CsakiErlichHollowood2000a,DvaliGabadadzePorrati2000a}. The GRS model also shed light on solving the cosmological constant problem~\cite{DvaliGabadadzePorrati2000a,Witten2000}. {But on the other hand, it suffers from problems such as violating the positivity of energy and containing ghost modes}~\cite{DvaliGabadadzePorrati2000a,Witten2000,CsakiErlichHollowood2000,GregoryRubakovSibiryakov2000a,PiloRattazziZaffaroni2000,CsakiErlichHollowoodTerning2001}.

{Another interesting and widely studied model which reproduces four-dimensional Newton's law at intermediate scale is the DGP model proposed by Dvali, Gabadadze, and Porrati~\cite{DvaliGabadadzePorrati2000}}. The DGP model offers us a new way to understand the acceleration of the universe~\cite{DeffayetDvaliGabadadze2002}. It is interesting to note that in higher-dimensional realization of the DGP model, the four-dimensional Newtonian gravity is reproduced by massive graviton resonances~\cite{KolanovicPorratiRombouts2003,GabadadzeShifman2004,RingevalRombouts2005}. In addition to the GRS and DGP models where the bulk is asymptotically flat, some works have considered graviton resonances in asymptotical AdS$_5$ space~\cite{ShaposhnikovTinyakovZuleta2004,Seahra2005,Seahra2005a,CveticRobnik2008}. Gravitational resonance also appears in other topic of gravity research. For instance, in black hole perturbation theory it is possible to find gravitational resonances which are known as the quasinormal modes~\cite{Nollert1999}. No doubt, gravitational resonances deserve further consideration.

In this paper we focus on the appearance of gravitational resonances in thick  RS-II type brane models, where the brane can be generated by, for example, a single canonical scalar field in an AdS$_5$ space~\cite{DeWolfeFreedmanGubserKarch2000,Gremm2000,KehagiasTamvakis2001,CsakiErlichHollowoodShirman2000} (more complicated thick brane models can be found in~\cite{DzhunushalievFolomeevMinamitsuji2010}).
The possibility of the existence of gravitational resonances in thick brane models was first pointed out in Ref.~\cite{Gremm2000}. But unfortunately, the solution therein does not support any gravitational resonance. In thick brane models constructed in general relativity, gravitational resonances can be found in cases with either a single canonical background scalar field~\cite{CruzGomesAlmeida2011,CruzMalufSousaAlmeida2016}, or two canonical scalar fields (the Bloch brane~\cite{BazeiaGomes2004})~\cite{CruzSousaMalufAlmeida2014,XieYangZhao2013}, or a single noncanonical scalar field (also known as K-field~\cite{Armendariz-PiconDamourMukhanov1999,GarrigaMukhanov1999,Armendariz-PiconMukhanovSteinhardt2001}; see~\cite{AdamGrandiSanchez-GuillenWereszczynski2008,BazeiaGomesLosanoMenezes2009,LiuZhongYang2010,BazeiaLobLosanoMenezes2013,ZhongLiuZhao2014a,BazeiaLobaoMenezes2015} for thick branes generated by K-fields)~\cite{ZhongLiuZhao2014}.

However, the spectrum of graviton is determined by the equation of the gravitational mode (the spin-2 transverse and traceless tensor mode of the metric perturbations~\cite{RandallSundrum1999a}). In general relativity, the equation of the gravitational mode is independent from the background scalar fields and only depends on the warp factor. But this is not always true in modified gravity theories. Thus, in order to find gravitational resonances, it is interesting to go beyond general relativity. For example, it is possible to find gravitational resonances in thick brane models in scalar-tensor gravity~\cite{GuoLiuZhaoChen2012}.

Another simply modified gravity is $f(R)$ gravity, where the Lagrangian of gravity is an arbitrary function of the scalar curvature. $f(R)$ gravity was first introduced in cosmology in 1970~\cite{Buchdahl1970}, and nowadays it has been applied in a wide range of cosmology studies~\cite{Starobinsky1980,Capozziello2002,NojiriOdintsov2003d,CarrollDuvvuriTroddenTurner2004,NojiriOdintsov2006a,AmendolaGannoujiPolarskiTsujikawa2007,HuSawicki2007,Starobinsky2007} (see~\cite{DeTsujikawa2010,SotiriouFaraoni2010,NojiriOdintsov2011} for reviews, and~\cite{NojiriOdintsov2000b,NojiriOdintsovOgushi2002a,ParryPichlerDeeg2005,AfonsoBazeiaMenezesPetrov2007,DeruelleSasakiSendouda2008,BalcerzakDabrowski2010,Bouhmadi-LopezCapozzielloCardone2010,DzhunushalievFolomeevKleihausKunz2010,HoffDias2011,LiuZhongZhaoLi2011,LiuLuWang2012,BazeiaMenezesPetrovSilva2013,BazeiaLobaoMenezesPetrovSilva2014,BazeiaLobaoLosanoMenezesOlmo2015,BazeiaLobaoMenezes2015} for $f(R)$ braneworld). The graviton equations for a class of thick $f(R)$-brane models were derived in Ref.~\cite{ZhongLiuYang2011}. It is shown that the graviton equation is determined by both the background solution and the function $f(R)$. Using the results of Ref.~\cite{ZhongLiuYang2011}, some of us find a series of gravitational resonances in Ref.~\cite{XuZhongYuLiu2015}, where the gravity Lagrangian density is $f(R)=R+\alpha R^2$, and the brane is generated by a single canonical scalar field. However, there exist some singularities in the model of~\cite{XuZhongYuLiu2015}.

The aim of this paper is to construct thick $f(R)$-brane models which support gravitational resonances but are free of singularities. By starting with some well-studied GR-brane solutions, we construct new $f(R)$-brane solutions (either analytically or numerically) such that the metric and the scalar configurations remain the same. Then we compare the gravitational resonance spectra between our $f(R)$-branes and their corresponding GR-branes. Since gravitons satisfy different equations in different gravity theories, there might be some differences in the resonance spectra.

Note that in addition to the tensor mode, there are also spin-1 vector modes and spin-0 scalar modes. These modes evolve independently, no matter the thick brane is constructed in general relativity~\cite{Giovannini2001a,Giovannini2002,KobayashiKoyamaSoda2002,Giovannini2003,ZhongLiu2013}, or in $f(R)$ gravity~\cite{ParryPichlerDeeg2005,ZhongLiu2015a}. The gravitational resonances only relate to the tensor sector. {Usually, the vector modes only contain an nonnormalizable zero mode and therefore can be neglected when one only considers static sources}~\cite{Giovannini2001a,Giovannini2002,ZhongLiu2013,ZhongLiu2015a}. The scalar sector will also contribute to the four-dimensional newtonian gravity~\cite{Giovannini2003}. However, for $f(R)$-brane generated by a background scalar field, the scalar sector of the linear perturbations has two degrees of freedom after fixing the gauge. In general, these two scalar modes are governed by two coupled second-order differential equations, which are too complicated to be considered here and deserve an independent work elsewhere.

In our first $f(R)$-brane model, we briefly discuss the similar model considered in Ref.~\cite{XuZhongYuLiu2015} and add the condition $f_R=\mathrm{d}f(R)/\mathrm{d}R>0$, which guarantees that the graviton is not a ghost and the normalization of the zero mode. In this model, we find that there is nothing special about gravitational resonances in $f(R)$-brane model in contrast to GR-brane. In order to get a different behavior, naturally, one can generalize this model to the case with multiple scalar fields. Based on this motivation, we study the brane generated by two scalar fields in second model and we obtain similar results as the case with a single scalar field. Our analysis shows that for the brane generated by canonical scalar fields (no matter how many they are), there is no striking difference between $f(R)$-brane and GR-brane in terms of gravitational resonance. Under the circumstances, noncanonical scalar fields may be a breakthrough. In our third $f(R)$-brane model, indeed, we find that even for the case with only one single noncanonical scalar field, gravitational resonances on $f(R)$-brane can be significantly different from the GR-brane.

Therefore our work is organized as follows. In Sect.~\ref{MainModel}, we review the tensor perturbations of $f(R)$-brane and the condition of normalizable zero mode. In sect.~\ref{sec3}, we construct some thick $f(R)$-brane models in cases with, first, a single canonical scalar, and then two canonical scalars, and finally a single noncanonical scalar field or K-field. These $f(R)$-brane models are all based on the corresponding GR-models which have been extensively studied in the literature. Then we compare, via both analytical and numerical methods, gravitational resonances between our $f(R)$-branes and the corresponding GR-branes. In Sect.~\ref{analyse}, we analyze how the $f(R)$-brane models impact gravitational resonances. Finally, in Sect.~\ref{Conclusion}, we come to the conclusions and discussions.

\section{Tensor perturbations of $f(R)$-brane}
\label{MainModel}

Now we consider the five-dimensional $f(R)$-brane model. Our action is taken in the general form
\begin{eqnarray}\label{action}
S=\int \mathrm{d}^4x \mathrm{d}y\sqrt{-g}\left(\frac{1}{2\kappa_5^2}f(R)
 +L(\phi_i,X_i)\right),
\end{eqnarray}
where $f(R)$ is an arbitrary function of the scalar curvature $R$ and $\kappa_5^2=8 \pi G_5$ with $G_5$  the five-dimensional Newton constant. For convenience, we take $\kappa_5^2=2$. Capital Latin indices $M, N, \ldots=0, 1, 2, 3, 4$ denote the bulk coordinates, and Greek indices $\mu,\nu, \ldots=0,1,2,3$ denote the brane coordinates. The coordinate $y=x^{4}$ stands for the extra dimension. $\phi_i$ is the $i$th scalar field, and its kinetic term $X_i$ is given by
\begin{eqnarray}
\label{X}
X_i=-\frac{1}{2}\partial^M\phi_i\partial_M\phi_i.
\end{eqnarray}

We consider a static flat brane, for which the line element can be written as
\begin{eqnarray}
\label{Metric1}
\mathrm{d}s^{2}=a^2(y)\eta_{\mu \nu }\mathrm{d}x^{\mu }\mathrm{d}x^{\nu
}+\mathrm{d}y^{2}, 
\end{eqnarray}
where $a^2(y)=e^{2A(y)}$ is the warp factor, and $\eta_{\mu\nu}$ is the induced metric on the brane. For this background space-time, the scalar fields only {depend} on the extra dimension, i.e., $\phi_{i}=\phi_{i}(y)$. The scalar curvature is given by
\begin{eqnarray}
R(y)=-4(5A'^2+2A''). \label{RyofA}
\end{eqnarray}
In this paper {the prime denotes the derivative} with respect to $y$. The Einstein equations and the equations of motion for the scalar fields are
{\begin{subequations}
\begin{eqnarray}
\label{EE1}
 f\!+\!2f_R\left(4A'^2 \!+\! A''\right)\!-\! 6f'_RA'\!-\!2f''_R\!+\!4L\!&=&\!0,\\
\label{EE2}
 f\!+\!8f_R\left(A''\!+\!A'^2\right)\!-\!8f'_RA'\!+\!4\sum_i L_{X_i} \phi'^2_i+4L\!&=&\!0,~~~~~~
\end{eqnarray}\label{Dy}
\end{subequations}}
and
\begin{eqnarray}
\label{Ephi}
\phi''_i(L_{X_i}\!+\!2X_i L_{X_iX_i})\!+\!L_{\phi_i}\!=\!2X_i L_{X_i\phi_i}\!-\!4L_{X_i}\phi'_i A',
\end{eqnarray}
respectively, where $L_{X_i}\equiv \partial L/\partial X_i$, $L_{\phi_i}\equiv \partial L/\partial \phi_i$, and $f_{R}\equiv
\mathrm{d}f(R)/\mathrm{d}R$.

{In this paper, we only consider the minimal coupling between $\phi_i$ and $X_i$.} Therefore, the Lagrangian density $L(\phi_i,X_i)$ is given by
\begin{eqnarray}
\label{L}
L(\phi_i,X_i)=\sum_i \big[F(X_i)-V(\phi_i)\big],
\end{eqnarray}
where $F(X_i)$ is an arbitrary function of $X_i$, and $V(\phi_i)$ is the scalar potential {of the scalar field $\phi_i$}.

Next, we consider the tensor perturbations of the background metric (\ref{Metric1}):
\begin{eqnarray}
\label{Metric2}
\mathrm{d}s^{2}=e^{2A(y)}(\eta_{\mu\nu}+h_{\mu\nu})\mathrm{d}x^{\mu }\mathrm{d}x^{\nu
}+\mathrm{d}y^{2},
\end{eqnarray}
where $h_{\mu\nu}$ satisfies the transverse and traceless (TT) conditions $\eta^{\mu\nu}h_{\mu\nu}=0=\partial_\mu h_\nu^\mu$.
With the coordinate transformation $\mathrm{d}z=a^{-1}\mathrm{d}y$, the perturbed Einstein equations read \cite{ZhongLiuYang2011}
\begin{eqnarray}
\label{SE}
\left[\partial_z^2+ \Big(3\frac{\partial_z a}{a}+\frac{\partial_z f_R}{f_R}\Big)\partial_z+\Box^{(4)}\right]
h_{\mu\nu}=0,
\end{eqnarray}
where $\Box^{(4)}=\eta^{\mu\nu}\partial_\mu\partial_\nu$. Note that {although in  Ref.~\cite{ZhongLiuYang2011} only one scalar field with canonical dynamics was considered, Eq.~(\ref{SE}) remains valid for our case because the TT tensor modes are decoupled from other modes of the perturbations.} Then, with the KK {decomposition \cite{ZhongLiuYang2011}}
\begin{eqnarray}
\label{decomposition}
h_{\mu\nu}(x^\rho,z)=\epsilon_{\mu\nu}(x^\rho) a^{-3/2}f_{R}^{-1/2}\psi(z),
\end{eqnarray}
where $\epsilon_{\mu\nu}(x^\rho)$ satisfies the TT conditions $\eta^{\mu\nu}\epsilon_{\mu\nu}=0=\partial_\mu\epsilon_\nu^\mu$, Eq.~(\ref{SE}) can be simplified as a Schr\"{o}dinger-like equation for $\psi(z)$ \cite{ZhongLiuYang2011}:
\begin{eqnarray}
\label{Schrodinger equation 1}
\left[-\partial_{z}^2+W(z)\right]\psi(z)=m^2\psi(z).
\end{eqnarray}
Here the effective potential $W(z)$ is given by \cite{ZhongLiuYang2011}
\begin{eqnarray}
\label{Wz}
W(z)\!\!&=&\!\!\frac34\frac{(\partial_{z}a)^2}{a^2}+\frac32\frac{\partial_z^{2}a}{a}+\frac32\frac{\partial_{z}a \partial_{z}f_R}{a f_R}\nonumber \\
  &-&\!\!\frac14\Big(\frac{\partial_z f_R}{f_R}\Big)^2+\frac12\frac{\partial_z^{2}f_R}{f_R}.
\end{eqnarray}
In the next section we will discuss the effective potential $W(z(y))$ in the coordinate $y$. Using the relation $\partial_z=a \,\partial_y$, Eq. (\ref{Wz}) can be rewritten in the coordinate $y$:
\begin{eqnarray}
\label{Wy}
W(z(y))&=&\frac{3}{4} a'^2+\frac{3}{2}(a'^2+aa'')+\frac{3aa'}{2}\frac{f'_R}{f_R}\nonumber\\
&-&  \frac{1}{4}\Big(\frac{a f'_R}{f_R}\Big)^2+\frac{1}{2}\frac{aa'f'_R+a^2f''_R}{f_R}.
\end{eqnarray}

From the Schr\"{o}dinger-like equation (\ref{Schrodinger equation 1}), it is easy to give the solution of the gravitational zero mode
\begin{eqnarray}
\label{zero mode}
\psi^{(0)}(z)=N_0a^{3/2}(z)f_{R}^{1/2}(z),
\end{eqnarray}
where $N_0$ is the normalization constant. To get four-dimensional Newtonian gravity on the brane, the zero mode should satisfy the following normalization condition:
\begin{eqnarray}
\int_{-\infty}^{\infty}|\psi^{(0)}(z)|^{2}\mathrm{d}z<\infty.
\end{eqnarray}
Apparently, in order to ensure the normalization of the zero mode, $f_R$ should satisfy $f_R>0$ in the whole space, which also guarantees that the graviton is not a ghost~\cite{SotiriouFaraoni2010}.

\section{Gravity resonances in various $f(R)$-brane models}
\label{sec3}
In this section we study braneworld solutions and gravitational resonances in three types of thick $f(R)$-brane models. In our first model, the brane is generated by a single canonical scalar field, which is one of the most widely studied thick brane models. Next, we consider the brane generated by two canonical scalar fields (known as the Bloch brane) in the second model. The main property of the Bloch brane is that the brane is split into two sub-branes. Finally, we study the brane generated by a K-field, {whose most important feature is} that its kinetic term is noncanonical, which will produce some different results compared to the previous two cases.

\subsection{$f(R)$-brane model with $L=X-V(\phi)$}
\label{sec3.1}

{First of all, we study the $f(R)$-brane generated by a single canonical scalar field, whose Lagrangian density reads} $L=X-V(\phi)=-\frac12\partial^M\phi\partial_M\phi-V(\phi)$.
{ This kind of $f(R)$-brane model is one of the simplest and the most widely studied models. Therefore, one reason why we study it is to review and illustrate the research processes of gravitational resonances on $f(R)$-brane. The second and also more important reason is to give a direct comparison with our third model generated by a single noncanonical scalar field.} The dynamical equations (\ref{Dy}) and (\ref{Ephi}) become
\begin{subequations}
\begin{eqnarray}
\label{a26}
 f\!+\!2f_R\left(4A'^2 \!+\! A''\right)
  \!-\! 6f'_RA'\!-\!2f''_R\!\!&=&\!\!2(\phi'^2\!+\!2V),~~~~~~   \\
\label{b26}
 -8f_R\left(A''\!+\!A'^2\right)\!+\!8f'_RA'
  -f\!\!&=&\!\!2(\phi'^2\!-\!2V), ~~~~~~  \\
\phi'' \!+\! 4 A' \phi' \!\!&=&\!\! V_\phi.
\end{eqnarray}
\end{subequations}

{For the case of $f(R)=R$, we consider the solution given in} \cite{Gremm2000}:
\begin{subequations}
\begin{eqnarray}
\phi(y) \!&=&\! \sqrt{6b} \arctan \left[\tanh\left(\frac{ky}{2}\right)\right],\label{phi11}\\
A(y)\!&=&\! -b\ln\left[\cosh (ky)\right],\label{warp1}\\
V(\phi) \!&=&\! \frac{3b k^2}{8}\left[(1-4b)+(1+4b)\cos \left(\sqrt{\frac{8}{3b}}\phi\right)\right],~~~~
\end{eqnarray}\label{GRBrane}
\end{subequations}
where $b$ and $k$ are positive parameters. For convenience, we set $\hat{y}=k y$ in the next content, which is a dimensionless parameter.

Now we construct the $f(R)$-brane model that shares the same background solution (\ref{phi11}) of {the GR-brane}.
Keeping the above background solution of the GR-brane unchanged, and incorporating Eqs.~(\ref{a26}) and (\ref{b26}), the function $f_R$ should satisfy
\begin{eqnarray}
\label{Cphi1}
 -\ddot{f}_R+\dot{f}_R \dot{A}-3f_R \ddot{A}=2\dot{\phi}^2\!\geq0,
\end{eqnarray}
where $\dot{A}=-b~\tanh(\hat{y})$, $\ddot{A}=-b~\text{sech}^2(\hat{y})$, $\dot{\phi}^2=\frac{3}{2} b~\text{sech}^2(\hat{y})$, and {the dot denotes the derivative} with respect to $\hat{y}$.

It is interesting to give {the} explicit expression for $f_R$ and $f(R)$. The {procedures are} described as follows. First, we solve Eq. (\ref{Cphi1}) to give the analytic solution for $f_R$ as a function of $\hat{y}$:
\begin{eqnarray}
 f_R\!\!&=&\!\!1\!+\!\alpha {\text{sech}^{\frac{b}{2}}(\hat{y})} \left[P_{K_{-}}^{\frac{b}{2}}\big(\tanh (\hat{y})\big)\!\!-\!\!  \beta
         Q_{K_{-}}^{\frac{b}{2}}\big(\tanh (\hat{y})\big)\right],\nonumber\\
 \label{f_R(y)}
\end{eqnarray}
where $\alpha$ is an arbitrary constant, $P$ and $Q$ are, respectively, associated Legendre functions of the first and second {kinds}, $\beta={P_{K_{+}}^{{b}/{2}}(0)}/{Q_{K_{+}}^{{b}/{2}}(0)}$ is given by the condition $\dot{f}_R|_{\hat{y}=0}=0$, and $K_{\pm}=\frac12\sqrt{(b-14) b+1}\pm1/2$.
Second, we define ${\hat{R}}\equiv \frac{R}{k^2}=4b \left[(5b+2)\text{sech}^2(\hat{y})-5b\right]$, which is obtained from Eqs.~(\ref{RyofA}) and (\ref{warp1}), to give the expression of $\hat{y}$ in terms of the curvature scalar ${\hat{R}}$:
\begin{eqnarray}
 \hat{y}=\text{arccosh}\left(\frac{2 \sqrt{2 b + 5 b^2}} {\sqrt{20b^2+{\hat{R}}}}\right). \label{y(R)}
\end{eqnarray}
Thirdly, defining $\hat{f}{\equiv}f/k^2$, which results in $\hat{f}_{\hat{R}}=f_R$, and substituting $\hat{y}({\hat{R}})$ into (\ref{f_R(y)}) yield the expression of $\hat{f}_{{\hat{R}}}$ as a function of ${\hat{R}}$. Finally, integrating $\hat{f}_{{\hat{R}}}$ with respect to ${\hat{R}}$ gives the expression of $\hat{f}({\hat{R}})$. In  the following, we will first consider some special values of the parameter $b$, which will result in simple solution of $\hat{f}(\hat{R})$, and then general $b$.

\subsubsection{$b=1$}

For the case of $b=1$, we obtain the following solution:
\begin{eqnarray}
 \hat{f}_{{\hat{R}}}=1+\alpha_1 \cos \left(\mathcal{H}({\hat{R}})\right), \label{f_R}
\end{eqnarray}
where $\mathcal{H}({\hat{R}})$ is given by
\begin{eqnarray}
\mathcal{H}({\hat{R}})= \sqrt{3} \ln \left(\frac{\sqrt{{{\hat{R}}}-8}+\sqrt{{{\hat{R}}}+20}}{2 \sqrt{7}}\right).
\end{eqnarray}
{Note that $\hat{f}_{{\hat{R}}}$ satisfies}  $\hat{f}_{{\hat{R}}}|_{\hat{y}=0}=1+\alpha_1$. Then the function $\hat{f}({\hat{R}})$ can be integrated from Eq.~(\ref{f_R}) as
\begin{eqnarray}
 \hat{f}({\hat{R}})\!\!&=&\!\!{\hat{R}}+ \frac{2\alpha_1}{7}\Big[ 2({\hat{R}}+6) \cos \left(\mathcal{H}({\hat{R}})\right)  \nonumber \\
     \!\!&+&\!\!  \sqrt{3({\hat{R}}-8)({\hat{R}}+20)} ~\sin \left(\mathcal{H}({\hat{R}})\right)
     \Big], \label{f}
\end{eqnarray}
where the integration constant is absorbed into the scalar potential $V(\phi)$.


\begin{figure*}
\begin{center}
\subfigure[$\alpha_1=0,100$]{\label{FR111}
\includegraphics[width=7cm,height=4.5cm]{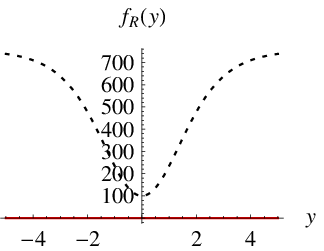}}
\subfigure[$\alpha_1=0,-0.13$]{\label{FR112}
\includegraphics[width=7cm,height=4.5cm]{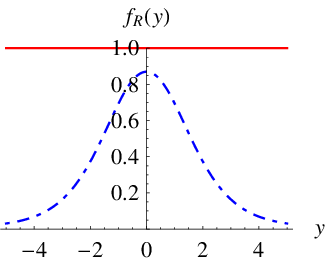}}
\subfigure[$\alpha_1=0,100,-0.13$]{\label{W11}
\includegraphics[width=7cm,height=4.5cm]{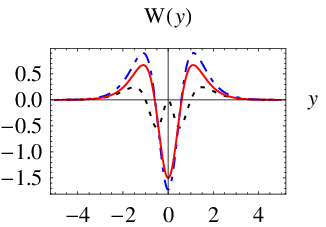}}
\subfigure[$\alpha_1=0,100,-0.13$]{\label{PM11}
\includegraphics[width=7cm,height=4.5cm]{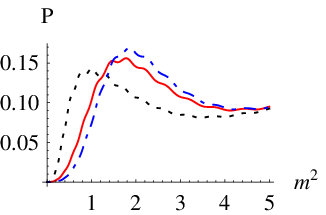}}
\end{center}
\caption{Plots of {the function} $f_R$, effective potential $W(z(y))$, and relative probability $P(m^2)$ of {the odd KK modes for} the single scalar brane with standard kinetic term {with $b=1$}. In the figures, the GR-brane ($\alpha_1=0$) is denoted by the red solid {lines}, {the $f(R)$-branes} with $\alpha_1=100$ and $\alpha_1=-0.13$ are donated by the black dotted lines and blue dot-dashed {lines}, {respectively}. (From now on, red solid lines always correspond to the GR-brane, black dotted {lines correspond} to $f_R>1$, blue dot-dashed lines correspond to $f_R<1$, and the rest of figures are denoted by black solid lines in all our {plots}).}\label{V1}
\end{figure*}

\subsubsection{$b=\frac{52}{3}$}

For the special case of $b=\frac{52}{3}$, we have
\begin{eqnarray}
 \hat{f}({{\hat{R}}})&=& {\hat{R}}+\alpha_2 \sum_{n=1}^{7} c_n \hat{R}^{n},\label{f2}
\end{eqnarray}
where the values of the coefficients $c_i$ are given in Table~\ref{table1}. The advantage of taking those values is that we can guarantee $\hat{f}_{{\hat{R}}}|_{\hat{y}=0}=1+\alpha_2$, which is similar to the case $b=1$.

\begin{table}
\begin{center}
\begin{tabular}{|c| c| }
\hline
~~$c_i$~~ &      Value \\
\hline
$c_1$ &      $\frac{240936850868416}{149442323051763}$\\
\hline
$c_2$ &      $-\frac{38202658243}{16604702561307}$ \\
\hline
$c_3$ &      $\frac{2065663433}{4605037510335808}$\\
\hline
$c_4$ &      $-\frac{18231247}{3831391208599392256}$\\
\hline
$c_5$ &      $-\frac{351657}{3187717485554694356992}$\\
\hline
$c_6$ &      $-\frac{6507}{2652180947981505705017344}$\\
\hline
$c_7$ &      $-\frac{459}{15446301841044289226021011456}$\\
\hline
\end{tabular}
\end{center}
\caption{The values of the parameters $c_i$ in Eq.~(\ref{f2}).}\label{table1}
\end{table}

\begin{figure*}
\begin{center}
\subfigure[$\alpha_2=0,100$]{\label{FR131}
\includegraphics[width=7cm,height=4.5cm]{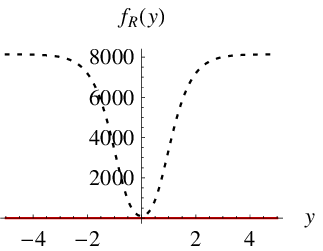}}
\subfigure[$\alpha_2=0,-0.012$]{\label{FR132}
\includegraphics[width=7cm,height=4.5cm]{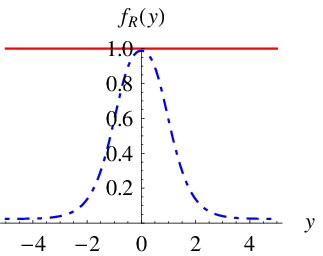}}
\subfigure[$\alpha_2=0,100,-0.012$]{\label{W13}
\includegraphics[width=7cm,height=4.5cm]{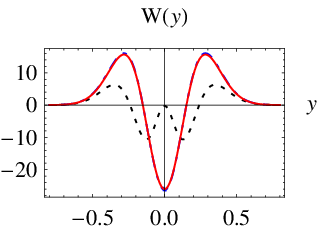}}
\subfigure[$\alpha_2=0,100,-0.012$]{\label{PM13}
\includegraphics[width=7cm,height=4.5cm]{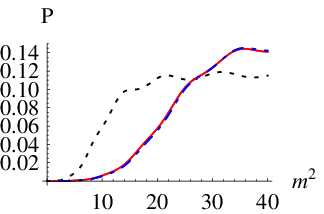}}
\end{center}
\caption{Plots of the function $f_R(y)$, effective potential $W(z(y))$, and relative probability $P(m^2)$ of the {odd KK modes for} the canonical single scalar brane with $b={52}/{3}$ and $\alpha_2=0,100,-0.012$, respectively. }\label{V3}
\end{figure*}

\subsubsection{The general $b$}

For general $b$, Eqs. (\ref{f_R(y)}) and (\ref{y(R)}) give the solution of $\hat{f}_{{\hat{R}}}$, which is a complex expression. In this case, we can give the analytic $\hat{f}({\hat{R}})$ from Eqs. (\ref{f_R(y)}), (\ref{y(R)}), and (\ref{a26}) or (\ref{b26}), which is given by
\begin{eqnarray}
 \hat{f}({{\hat{R}}})
 \!\!&=&\!\!    \hat{R}
    +\!\alpha \bigg\{ \frac{24b^2\!+\!2{\hat{R}}\!+ \!2b{\hat{R}}}{2\!+\!5b}
     \Big[P_{K_{-}}^{{b}/{2}}(\Psi)
    \!-\!   {\beta}Q_{K_{-}}^{{b}/{2}}(\Psi)
     \Big]  \nonumber \\
    &-&\!\!
 4(b^2\!-\!2bK_{+}) \Psi\Big[P_{K_{+}}^{{b}/{2}}(\Psi)
     -\!\! \Psi P_{K_{-}}^{{b}/{2}}(\Psi) \nonumber \\
    &+&\!\!\beta \Psi\left( Q_{K_{-}}^{{b}/{2}}(\Psi)
    - Q_{K_{+}}^{{b}/{2}}(\Psi)\right)\Big] \bigg\}  \Theta^{b/2}, \label{fR_b}
\end{eqnarray}
where the parameters $\alpha$, $\beta$, and $K_{\pm}$ {are those appearing in (\ref{f_R(y)})} and $\Theta$ and $\Psi$ are given by
\begin{eqnarray}
\Psi&=&\sqrt{1-\Theta^2},\\
\Theta&=&\frac{\sqrt{20b^2+{\hat{R}}}}{2\sqrt{2b+5b^2}}.
\end{eqnarray}

To investigate the resonant modes of gravity we recall the definition of the relative probability~\cite{LiuYangZhaoFuDuan2009}:
\begin{eqnarray}
P(m^{2})=\frac{\int^{z_{b}}_{-z_{b}}|\psi(z)|^{2}dz}{\int^{z_{max}}_{-z_{max}}|\psi(z)|^{2}dz},
\end{eqnarray}
where $\psi(z)$ is the solution of Eq.~(\ref{Schrodinger equation 1}), $2z_b$ is approximately the width of the brane, and $z_\mathrm{max}=10z_b$. Here $|\psi(z)|^{2}$ can be explained as the probability density~\cite{LiuYangZhaoFuDuan2009,AlmeidaFerreiraGomesCasana2009}. A resonant mode with mass $m_n$ exists, if the relative probability $P(m^2)$ has a peak around $m=m_n$ and this peak has a full width at half maximum (FWHM). So, the number of the peaks that have FWHM corresponds to the number of the resonant modes. For the sake of simplicity, we impose the following conditions on $\psi(z)$:
\begin{subequations}
\begin{eqnarray}
\label{even}
\psi_{\rm{even}}(0)\!\!&=&\!\!1, ~~~\partial_{z}\psi_{\rm{even}}(0)=0;\\
\label{odd}
\psi_{\rm{odd}}(0)\!\!&=&\!\!0, ~~~~\partial_{z}\psi_{\rm{odd}}(0)=1,
\end{eqnarray}\label{EvenOddConditions}
\end{subequations}
where $\psi_{\rm{even}}$ and $\psi_{\rm{odd}}$ correspond to the even and odd parity modes of $\psi(z)$, respectively. Since there is no essential difference between the above conditions for the relative probability $P(m^2)$, we will only display our results with one of the conditions (\ref{EvenOddConditions}) in the following examples. For the sake of brevity, we discuss gravitational resonance in the coordinate $\hat{y}$, but set $y=\hat{y}$, $R={\hat{R}}$ and $f={\hat{f}}$, which is {equivalent} to setting ${k=1}$ in the coordinate $y$.

{Note that in this $f(R)$-brane model there are two primary parameters $b$ and $\alpha$. The parameter $b$ determines the brane solution and $\alpha$ affects the range of values of the function $f(R)$. Therefore, they may characterize the gravitational resonances that will be discussed in the following. It can be seen that the expression of $f(R)$ in (\ref{fR_b}) for general $b$ is very complex while it becomes simple for two special values: $b=1$ and $b=52/3$. We find that the shapes of $f(R(y))$ as the functions of $y$ for the two special and general values of $b$ are similar. Therefore, in the following calculations, we only need consider the two special values, for which the expressions of $f(R)$ and the effective potential $W(y)$ are simple. As for the parameter $\alpha$, it characterizes the deviation of the $f(R)$-brane from the GR-brane. This can be seen from (\ref{f_R(y)}).}

For {the case of} $b=1$, we substitute (\ref{f}) into (\ref{Wy}) to solve Schr\"{o}dinger-like Eq~(\ref{Schrodinger equation 1}) {with three typical values of $\alpha_1$}:  $\alpha_1=0$ (GR-brane), $\alpha_1=100$ ($f(R)$-brane with $f_R>1$), and $\alpha_1=-0.13$ ($f(R)$-brane with $0<f_R<1$) (see Fig.~\ref{FR111}, \ref{FR112}). Note that the functions $f_R(y)$ for $\alpha_1=100$ and $\alpha_1=-0.13$ (both guarantee $f_R(y)>0$ in the whole space) have already deviated from $f_R(y)=1$ for $\alpha_1=0$ (the GR-brane) as far as possible. For $\alpha>100$, the potential $W(z(y))$ can hardly change more compared with $\alpha_1=100$. For $\alpha_1<-0.13$, $f_R$ {may be negative} at some points. The effective potentials $W(z(y))$ are shown in Fig.~\ref{W11}. Our calculation shows that for all the chosen values of $\alpha_1$ (see Fig.~\ref{PM11}), there is no gravitational resonance.

For the case of $b=\frac{52}{3}$, the corresponding solution of $f(R)$ is given by Eq.~(\ref{f2}). Compared with {the case of $b=1$, although the maximum of the potential $W(z(y))$ is enlarged, but no resonant KK mode is found}. The corresponding results are shown in Fig.~\ref{V3}. For other values of $b$, the results are similar: the $f(R)$-brane can hardly deviate from the corresponding GR-brane in terms of gravitational resonance and {we do not find any gravitational resonance.}

\subsection{$f(R)$-Bloch brane model with $L=X_1+X_2-V(\phi_1,\phi_2)$}
\label{sec3.2}

{Next let us discuss the Bloch brane, which is a kind of important brane with inner structure generated by two scalar fields. It was first obtained in Ref.~\cite{BazeiaGomes2004} and then developed and discussed in Refs.}~\cite{SouzaFariaHott2008,XieYangZhao2013,CruzMalufAlmeida2013,CruzLimaAlmeida2013,SouzaBritoHoff2014,ZhaoLiuZhong2014,XieGuoZhaoDuZhang2015}.
{ In this subsection, we explore gravitational resonances on the Bloch brane in the $f(R)$ gravity (called the $f(R)$-Bloch brane) for several reasons. First, there exist gravitational resonances on the Bloch brane~\cite{XieYangZhao2013}. Therefore, we can compare the difference between the GR-Bloch brane and $f(R)$-Bloch brane in the presence of gravitational resonances, which is expected to be a complement to the former $f(R)$-brane model. Second, the Bloch brane is generated by two canonical scalar fields. Investigation of the Bloch brane in this subsection could pave the way for Sect.~\ref{analyse}, where we will generalize the conclusion of the first two models to the cases with multiple canonical scalar fields.}

The Lagrangian density~(\ref{L}) for the Bloch brane is given by $L=-\frac{1}{2}\partial^M\phi\partial_M\phi-\frac{1}{2}\partial^M\chi\partial_M\chi-V(\phi,\chi)$, where $\phi$ and $\chi$ are two interacted real scalar fields.

The dynamical equations (\ref{Dy}) and (\ref{Ephi}) are modified to
\begin{subequations}
\begin{eqnarray}
\label{a27}
 f\!+\!2f_R\left(4A'^2 \!+\! A''\right)
  \!-\! 6f'_RA'\!-\!2f''_R\!\!\!&=&\!\!\!2(\phi'^2\!+\chi'^2\!+\!2V),~~~~~~   \\
\label{b27}
 -\! f-8f_R\left(A''+A'^2\right)+8f'_RA'
  \!\!\!&=&\!\!\!2(\phi'^2\!+\chi'^2\!-\!2V), ~~~~~~  \\
\phi'' + 4 A'\phi' & = & V_\phi, \\
\chi'' + 4 A'\chi' & = & V_\chi.
\end{eqnarray}
\end{subequations}

{For $f(R)=R$, there are many symmetric and asymmetric solutions}~\cite{BazeiaGomes2004,SouzaFariaHott2008,SouzaFariaHott2008}. We consider {one of the solutions} given in Refs.~\cite{BazeiaGomes2004,SouzaFariaHott2008}, where the scalar potential $V(\phi)$ takes the following form:
\begin{eqnarray}
V(\phi) &=& \frac12 \left[\left(\tilde{b} v^2-\tilde{b}\phi^2-d\chi^2\right)^2+4d^2 \phi^2 \chi^2 \right]\nonumber \\
        &-&\frac43\left(\tilde{b}\phi v^2-\frac13 \tilde{b}\phi^3-d\phi\chi^2\right)^2.
\end{eqnarray}
For the case of $\tilde{b}>2d>0$, the general symmetric Bloch-brane solution reads~\cite{BazeiaGomes2004,SouzaFariaHott2008}
\begin{subequations}
\begin{eqnarray}
\label{BlochBrane1Phiy}
 \phi(y) \!\!&=&\!\! v \tanh (2dvy),\\
\label{BlochBrane1Chiy}
 \chi(y) \!\!&=&\!\!v \sqrt{\frac{\tilde{b}\!-\!2d}{d}}~\text{sech} (2dvy),\\
\label{BlochBrane1Ay}
 A(y) \!\!&=&\!\! \frac{v^2}{9d}
    \left[ (\tilde{b}\!-\!3d)\tanh^2 (2dvy)
           \!-\!2\tilde{b} \ln \cosh(2dvy) \right].~~~~~~
\end{eqnarray}
\label{BlochBrane1}
\end{subequations}

Analogously, selecting certain {values} of {the parameters} $v$ and $\tilde{b}$, we can calculate {the} explicit expression of $f(R)$. For $v=\sqrt{3/2}$ and $\tilde{b}=3d$, we have
\begin{eqnarray}
f_R=1+\gamma \cos \left(\mathcal{Y}(R)\right),
\end{eqnarray}
where $\gamma$ is a constant, $R=24 d^2 \left[7~\text{sech}^2\left(\sqrt{6} \mathrm{d} y\right)-5\right]$, and $\mathcal{Y}(R)$ reads
\begin{eqnarray}
\mathcal{Y}(R)=\sqrt{3}\ln \left(\frac{\sqrt{R-48 d^2}+\sqrt{120 d^2+R}}{2 \sqrt{42}d}\right).
\end{eqnarray}
Therefore, integrating the function $f_{R}$ over $R$ we get
\begin{eqnarray}
f(R)\!\!&=&\!\!R+ \frac{2\gamma}{7}\Big[\sqrt{3(R-48 d^2)(120 d^2+R)} \sin \mathcal{Y}(R)\nonumber \\
     \!\!&+&\!\! 2\left(36 d^2+R\right) \cos \mathcal{Y}(R)\Big].
\end{eqnarray}
Note that when $d=\sqrt{\frac{1}{6}}$, this solution is identical with Eq.~(\ref{f}).

\begin{figure*}
\begin{center}
\subfigure[$f_R|_{y=0}=1,50$]{\label{FR21}
\includegraphics[width=5.5cm,height=4.5cm]{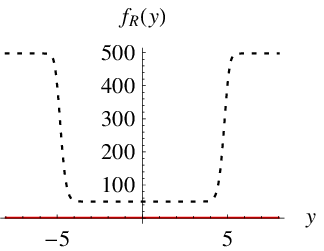}}
\subfigure[$f_R|_{y=0}=1,0.92$]{\label{FR22}
\includegraphics[width=5.5cm,height=4.5cm]{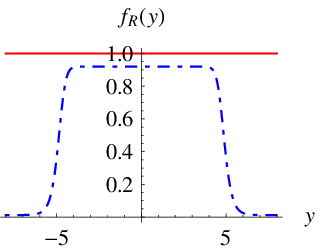}}
\subfigure[$f_R|_{y=0}=1,50,0.92$]{\label{W21}
\includegraphics[width=5.5cm,height=4.5cm]{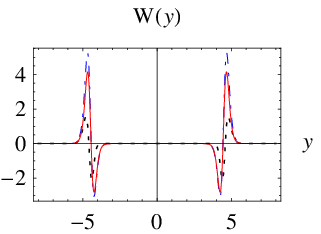}}
\subfigure[$f_R|_{y=0}=1,50,0.92$]{\label{PM2}
\includegraphics[width=0.6\textwidth]{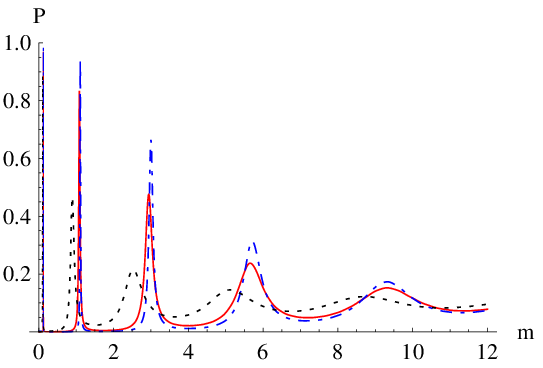}}
\end{center}
\caption{Plots of the {function} ${f_R}(y)$, effective potential $W(z(y))$, and relative probability $P(m^2)$ of odd resonance KK modes on the Bloch brane {with $d=1, v=1, c_0=-2-7.0\times 10^{-16}$, and $f_R|_{y=0}=1,50,0.92$, respectively}.}
\end{figure*}

For the case of $\tilde{b}=d$, the authors of Ref.~\cite{SouzaFariaHott2008} obtained the degenerate Bloch-brane solution:
\begin{subequations}
\begin{eqnarray}
\label{DBbrane1phi}
 {\phi}(y)
   \!\!&=&\!\!  v\frac{\sqrt{c_{0}^{2}-4}\sinh (2dvy)}
              {\sqrt{c_{0}^{2}-4}\cosh (2dvy)-c_{0}},
              \\
\label{DBbrane1chi}
 {\chi}(y)
   \!\!&=&\!\! \frac{2v}{\sqrt{c_{0}^{2}-4}\cosh (2dvy)-c_{0}},
               \\
\label{DBbrane1E2A}
 e^{2A(y)}
  \!\!&=&\!\! \left( \frac{\sqrt{c_{0}^{2}-4}-c_{0}}
         {\sqrt{c_{0}^{2}-4}\cosh (2dvy)-c_{0}}\right)
          ^{4v^{2}/9} \nonumber \\
  &\times&\!\!\exp \left[ -\frac{4v^2\left( c_{0}^{2}-4-c_{0}
     \sqrt{c_{0}^{2}-4}\right) }
     {9\left( \sqrt{c_{0}^{2}-4}
     -c_{0}\right) ^{2}}\right]\nonumber \\
  &\times&\!\! \exp \left[ \frac{4v^2\left( c_{0}^{2}\!-\!4\!-\!c_{0}
     \sqrt{c_{0}^{2}\!-\!4}\cosh(2dvy)\right) }
     {9\left( \sqrt{c_{0}^{2}\!-\!4}\cosh(2dvy)
     \!-\!c_{0}\right) ^{2}}\right],~~~~~~~
\end{eqnarray}\label{DBbrane1}
\end{subequations}
where $c_{0}<-2$.

Next, we take the above Bloch-brane solution (\ref{DBbrane1}) as our background solution of $f(R)$-brane. Using the similar process used in the previous section, the {function} $f_R$ should satisfy
\begin{eqnarray}
 -f''_R+f'_R A'-3f_R A''=2(\phi'^2\!+\chi'^2\!)\geq0.\label{a28}
\end{eqnarray}
Since it is difficult to {give the} analytical {solution} of $f_R$, we can obtain $f_R$ (also $f(R)$) and the potential $W(z(y))$ by means of numerical methods. Because $f(R(y))$ (and hence $f_R$) is an even function of the extra dimension $y$, we have $f'_R|_{y=0}=0$. To construct a numerical $f_R$ satisfying Eq.~(\ref{a28}), we need one more initial condition, i.e., the value of $f_R|_{y=0}$. With the adjustable parameter $f_R|_{y=0}$, we can construct a different numerical $f_R$.

{Here, we consider the following typical set of parameters reducing a double-brane: $a=d=1,~v=2$, and $c_0=-2\)¨C\(7.0\times 10^{-16}$, which was used in Ref.~\cite{XieYangZhao2013} for the GR-brane model. It is convenient to compare the results of the $f(R)$-brane and the GR-brane.}

The effective potential $W(z(y))$ and the relative probability $P(m^2)$ of the gravitational KK modes are shown, respectively, in {Fig.~\ref{W21}, \ref{PM2}}. We find that there are five resonance peaks in Fig.~\ref{PM2} for the GR-brane. {For the $f(R)$-brane model, the potential $W(z(y))$ (see Fig.~\ref{W21}) almost reaches the maximum deviation from the GR-brane when $f_R|_{y=0}=50$ and 0.92 (the numerical solutions of $f_R$ are shown in Fig.~\ref{FR21}, \ref{FR22}). The corresponding resonance mass spectra are shown in Fig.~\ref{PM2}, from which we can see that there exist four peaks for $f_R|_{y=0}=50$ and five peaks for $f_R|_{y=0}=0.92$, respectively. Therefore, there is no obvious difference between these spectra.} {For other values of the parameters we also have the same conclusion.}

\subsection{$f(R)$-brane model with $L=X-\lambda X^2-V(\phi)$}
\label{sec3.3}

{In the previous two subsections, the $f(R)$-branes are all generated by one or two canonical scalar fields. For comparison, we try to investigate $f(R)$-brane generated by noncanonical scalar fields (K-fields) in this subsection.}

{ K-fileds were introduced to study inflation for the first time~\cite{Armendariz-PiconDamourMukhanov1999,GarrigaMukhanov1999,Armendariz-PiconMukhanovSteinhardt2001}
and then were extended to many fields. In brane world models, branes generated by K-fields is generally known as K-branes, which might present some new properties (such as the localization of bulk fermions and gravitons~\cite{ZhongLiuZhao2014a}) compared with the corresponding standard branes. Therefore, we expect there are some new results of gravitational resonances on K-branes.}

{
We consider the simple case of one K-field with the Lagrangian density $L=X-\lambda X^2-V(\phi)$ ($\lambda\neq0$). The dynamical equations (\ref{Dy}) and (\ref{Ephi}) are reduced to}
\begin{eqnarray}
\label{EE31}
 f\!\!+\!\!2f_R\left(4A'^2 \!\!+\!\! A''\right)
  \!\!-\!\! 6f'_RA'\!\!-\!\!2f''_R\!\!&=&\!\!2\left(\phi'^2\!\!+\!\!\frac{1}{2}\lambda \phi'^4\!\!+\!\!2V\right),~~~~~~\nonumber \\
\label{EE32}
 -f\!\!-8f_R\left(A''\!\!+\!\!A'^2\right)+8f'_RA'\!\!&=&\!\!2\left(\phi'^2\!\!+\!\!\frac{3}{2}\lambda \phi'^4\!\!-\!\!2V\right),~~~~~~\\
\label{Ephi3}
  \phi''\!\!+\!\!4A'\phi'\!\!+\!\!\lambda(3\phi''\!\!+\!\!4A'\phi')\phi'^2\!\!&=&\!\!V_{\phi}.\nonumber
\end{eqnarray}

\begin{figure*}
\begin{center}
\subfigure[$b=2,~d=2$]{\label{1phi1}
\includegraphics[width=5cm,height=4cm]{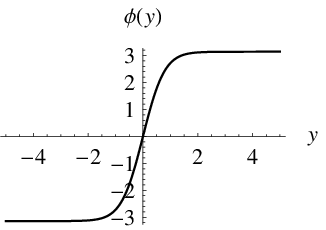}}
\subfigure[$b=1,~d=-2$]{\label{1phi2}
\includegraphics[width=5cm,height=4cm]{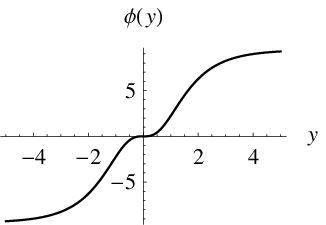}}
\subfigure[$b=6,~d=-6$]{\label{1phi3}
\includegraphics[width=5cm,height=4cm]{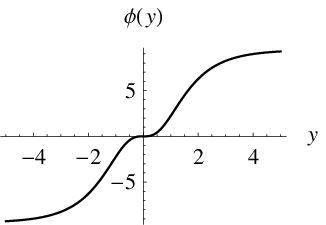}}
\end{center}
\caption{Plots of the scalar field $\phi$ given by Eq.~(\ref{0phi1}). The figures are, from left to right, kink-like solution, non-monotonic {kink} solution, and double kink-like solution.}\label{PHI}
\end{figure*}

For $f(R)=R$, the authors in Ref.~\cite{ZhongLiuZhao2014a} obtained one analytic solution. For general $f(R)$, in the vast majority of cases, the dynamical equations remain hard to solve except in some specific cases (for example, if $f(R)$ is a polynomial). As an example, we consider the following scalar field and warp factor:
\begin{eqnarray}
\phi(y) \!&=&\! \omega \arctan \left[\tanh\Big(\frac{y}{2}\Big)\right],\\
A(y)\!&=&\!-\ln\left[\cosh (y)\right].
\end{eqnarray}
For the case of $\omega=1$, we get the solution
\begin{eqnarray}
f(R)=R+\frac{\lambda}{3136} R^2+\frac{(29 \lambda -980)}{1176} R,
\end{eqnarray}
where $R=28~\text{sech}^2(y)-20$ and $\lambda>-\frac{10976}{1955}$ to guarantee $f_R>0$.

\begin{figure*}
\begin{center}
\subfigure[$f_R|_{y=0}=1,100$]{\label{W31}
\includegraphics[width=7cm,height=4.5cm]{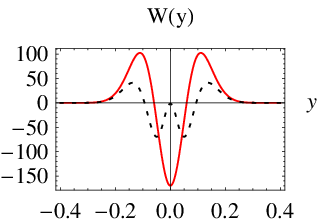}}
\subfigure[$f_R|_{y=0}=1,100$]{\label{PM31}
\includegraphics[width=7cm,height=4.5cm]{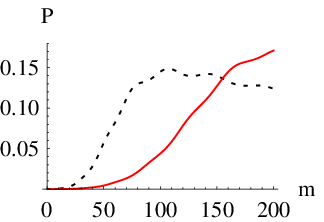}}
\end{center}
\caption{Plots of the effective potential $W(z(y))$ and relative probability $P(m^2)$ of {the} odd resonance KK modes on the brane generated by a K-field with $b=2,d=2,\lambda=0.6$, and $f_R|_{y=0}=1,100$.}\label{ALLPM31}
\end{figure*}

\begin{figure*}
\begin{center}
\subfigure[$f_R|_{y=0}=1,100$]{\label{W32}
\includegraphics[width=7cm,height=4.5cm]{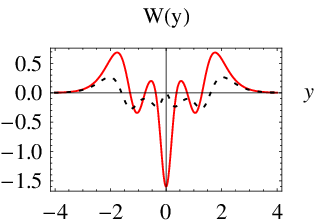}}
\subfigure[$f_R|_{y=0}=1,100$]{\label{PM32}
\includegraphics[width=7cm,height=4.5cm]{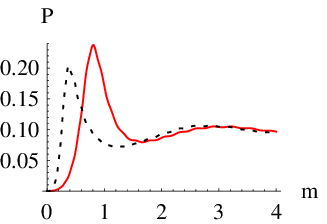}}
\end{center}
\caption{Plots of the effective potential $W(z(y))$ and relative probability $P(m^2)$ of odd resonance KK modes on the brane generated by a K-field with $b=1,d=-2,\lambda=0.6$, and $f_R|_{y=0}=1,100$.}\label{ALLPM32}
\end{figure*}

\begin{figure*}
\begin{center}
\subfigure[$f_R|_{y=0}=1,100$]{\label{W33}
\includegraphics[width=7cm,height=4.5cm]{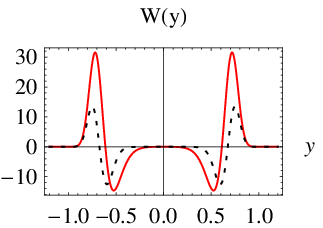}}
\subfigure[$f_R|_{y=0}=1,100$]{\label{PM33}
\includegraphics[width=7cm,height=4.5cm]{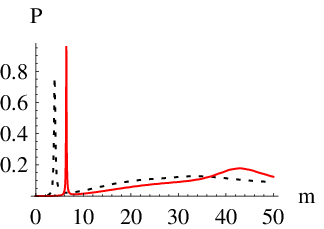}}
\end{center}
\caption{Plots of the effective potential $W(z(y))$ and relative probability $P(m^2)$ of the odd resonance KK modes on the brane generated by K-field {with $b=6,d=-6,\lambda=0.6$, and $f_R|_{y=0}=1,100$}.}\label{ALLPM33}
\end{figure*}

If we require the coefficient of $R$ to be 1, then we obtain
\begin{eqnarray}
\phi(y) \!\!&=&\!\!  \sqrt{\frac{\pm14\sqrt{174 \lambda\!+\!49}\!-\!98}{29\lambda }}
     \arctan \left[\tanh\Big(\frac{y}{2}\Big)\right],\label{42}\\
A(y) \!\!&=&\!\! -\ln\left[\cosh (y)\right],\\
f(R) \!\!&=&\!\! R+\frac{\left(87 \lambda \mp7 \sqrt{174 \lambda +49}+49\right)}{6728 \lambda } R^2.
\end{eqnarray}
For $+$ in Eq.~(\ref{42}), we require $-\frac{49}{174}<\lambda<\frac{5}{2}$ and $-$ corresponds to $-\frac{49}{174}<\lambda<-\frac{4}{25}$. This solution {was} also obtained in Ref.~\cite{BazeiaLobaoLosanoMenezesOlmo2015} {recently with another method}.

Here, to study the solution of $\phi$ with more abundant structure, we reconstruct one analytic solution of $\phi$ given as
\begin{eqnarray}
\label{0phi1}
\phi=b \arcsin\left[\tanh(y)\right]+d \tanh(y) \text{sech}(y),
\end{eqnarray}
where $b$ and $d$ are free parameters. Therefore, the warp factor is {determined} by
\begin{eqnarray}\label{0phi2}
 2\lambda\phi'^4+3A'' +2\phi'^2=0,
\end{eqnarray}
{with the boundary conditions} $A(0)=0$ and $A'(0)=0$.
Meanwhile, $f_R$ should satisfy
\begin{eqnarray}
\label{fR3}
2\lambda\phi'^4+3f_R A''+f_R''-f_R'A'+2\phi'^2=0.
\end{eqnarray}
The value of $f_R|_{y=0}$ can be arbitrary so long as the corresponding $f_R$ satisfies Eq.~(\ref{fR3}) and guarantees $f_R>0$ in the whole space.

If the parameter $\lambda$ is positive, the solution of {the} wrap factor will be lumplike and the constraint condition (\ref{fR3}) will be the same as Eqs.~(\ref{Cphi1}) and (\ref{a28}). Our results show that it has no breakthrough comparing with the former cases {discussed in the previous two subsections}.

For $\lambda>0$, we consider three types of solutions of the scalar field (see Fig.~\ref{PHI}). Because when $0<f_R|_{y=0}<1$ the range for $f_R|_{y=0}$ is so narrow, we only consider the case $f_R|_{y=0}\geq1$ in this $f(R)$-brane model. {We find that when the scalar field $\phi$ is taken as the form of single kink or non-monotonic kink, the results are the same as the brane generated by a canonical scalar field (see Figs.~\ref{ALLPM31}, \ref{ALLPM32}). For the case of double kink, the result is similar to the Bloch brane (see Fig.~\ref{ALLPM33}).}

\begin{figure*}
\begin{center}
\subfigure[$b=1,d=0.3$]{\label{PHI1}
\includegraphics[width=5cm,height=4cm]{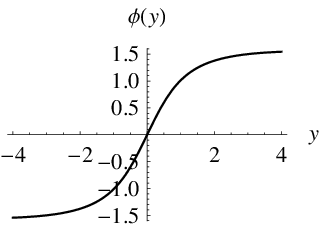}}
\subfigure[$b=1,d=-2$]{\label{PHI2}
\includegraphics[width=5cm,height=4cm]{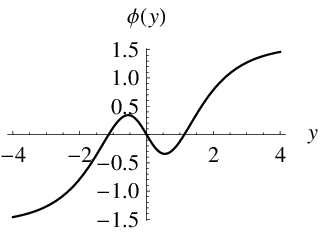}}
\subfigure[$b=8,d=-8$]{\label{PHI2}
\includegraphics[width=5cm,height=4cm]{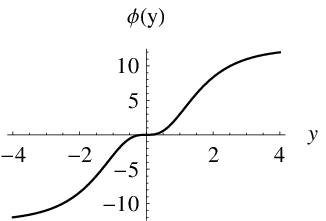}}
\end{center}
\caption{Plots of the scalar field $\phi$ given by Eq.~(\ref{0phi1}). The figures are, from left to right, a kink-like solution, a non-monotonic kink solution, and a double kink-like solution.}\label{9a}
\end{figure*}

\begin{figure*}
\begin{center}
\subfigure[$f_R|_{y=0}=1,50$]{\label{W41}
\includegraphics[width=7cm,height=4.5cm]{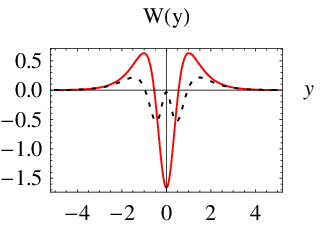}}
\subfigure[$f_R|_{y=0}=1,50$]{\label{PM41}
\includegraphics[width=7cm,height=4.5cm]{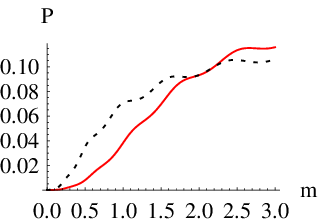}}
\end{center}
\caption{Plots of the effective potential $W(z(y))$ and relative probability $P(m^2)$ of the odd resonance KK modes on the brane generated by K-field with $b=1,d=0.3,\lambda=-0.01$, and $f_R|_{y=0}=1,50$.}\label{ALLW41}
\end{figure*}

\begin{figure*}
\begin{center}
\subfigure[$f_R|_{y=0}=1,50$]{\label{W42}
\includegraphics[width=7cm,height=4.5cm]{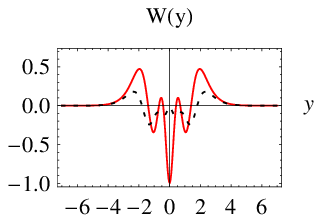}}
\subfigure[$f_R|_{y=0}=1,50$]{\label{PM42}
\includegraphics[width=7cm,height=4.5cm]{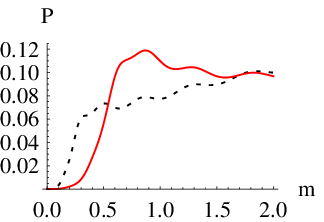}}
\end{center}
\caption{Plots of the effective potential $W(z(y))$ and relative probability $P(m^2)$ of the odd resonance KK modes on the brane generated by K-field with $b=1,d=-2,\lambda=-0.01$, and $f_R|_{y=0}=1,50$.}\label{ALLW42}
\end{figure*}

\begin{figure*}
\begin{center}
\subfigure[$f_R|_{y=0}=1,50$]{\label{W43}
\includegraphics[width=7cm,height=4.5cm]{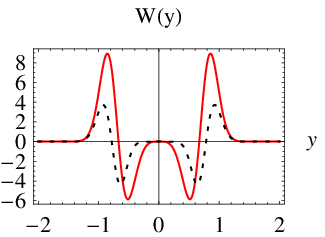}}
\subfigure[$f_R|_{y=0}=1,50$]{\label{PM43}
\includegraphics[width=7cm,height=4.5cm]{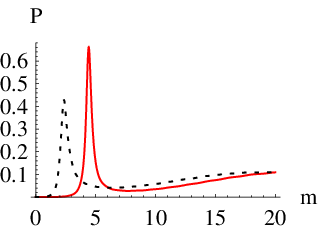}}
\end{center}
\caption{Plots of the effective potential $W(z(y))$ and relative probability $P(m^2)$ of the odd resonance KK modes on the brane generated by K-field with $b=8,d=-8,\lambda=-0.01$, and $f_R|_{y=0}=1,50$.}\label{ALLW43}
\end{figure*}

For $\lambda<0$, we also set three similar groups of parameters (see Fig.~\ref{9a}) and all of them have lumplike solutions for warp factor. Here we also only consider the values of parameters which result in $f_R|_{y=0}\geq1$ in  $f(R)$-brane model. Obviously, the consequences (see Figs.~\ref{ALLW41},~\ref{ALLW42}, \ref{ALLW43}) are analogous to the case $\lambda>0$.

It seems that we can draw the conclusion that the difference of the gravitational resonances is small between the $f(R)$-brane and GR-brane when the warp factor is monotonic in the area $ y\in [0,+\infty)$.

\begin{figure*}
\begin{center}
{\includegraphics[width=7cm,height=4.5cm]{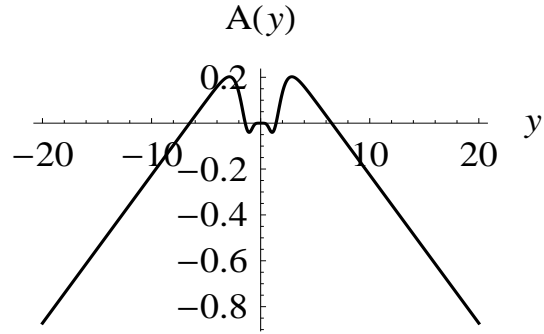}}
\end{center}
\caption{{Plot} of the warp factor $A(y)$ given by Eq.~(\ref{0phi2}) with $b=2,d=-2$, and $\lambda=-0.6$. The parameters correspond to a double kink-like solution of {the} scalar field.}\label{nonmonotonicA}
\end{figure*}

\begin{figure*}
\begin{center}
\subfigure[$f_R|_{y=0}=1$]{\label{W441}
\includegraphics[width=7cm,height=4.5cm]{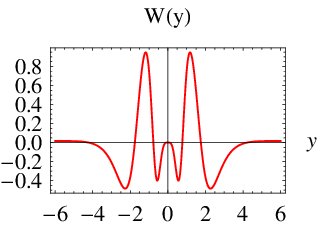}}
\subfigure[$f_R|_{y=0}=0.11$]{\label{W442}
\includegraphics[width=7cm,height=4.5cm]{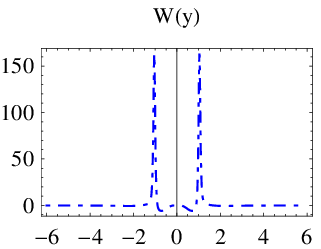}}
\subfigure[$f_R|_{y=0}=1,0.11$]{\label{PM44}
\includegraphics[width=9cm,height=4.5cm]{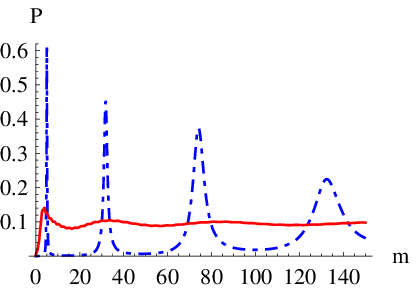}}
\end{center}
\caption{Plots of the effective potential $W(z(y))$ and relative probability $P(m^2)$ of the odd resonance KK modes on the brane generated by K-field with $b=2,d=-2,\lambda=-0.6$, and $f_R|_{y=0}=1,0.11$.}
\end{figure*}

Therefore, we consider another interesting scenario, i.e., the warp factor is a non-monotonic function {of $y$} in the area $ y\in [0,+\infty)$. {The} parameters are taken as $b=2,~d=-2$, and $\lambda=-0.6$ {(see Fig.~\ref{nonmonotonicA})}. The effective potential $W(z(y))$ and relative probability $P(m^2)$ of the GR-brane are shown in Fig.~\ref{W441}, \ref{PM44}. Apparently, there is no gravitational resonance on the GR-brane. Setting the same parameters for {the} $f(R)$-brane model, it can be seen that when $f_R|_{y=0}=0.11$, in comparison to the GR-brane the effective potential and relative probability change greatly (see Fig.~\ref{W442}, \ref{PM44}).

\begin{figure*}
\begin{center}
\subfigure[$f_R|_{y=0}=0.106$]{\label{FR55}
\includegraphics[width=7cm,height=4.5cm]{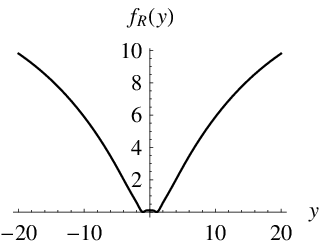}}
\subfigure[$f_R|_{y=0}=0.106$]{\label{W55}
\includegraphics[width=7cm,height=4.5cm]{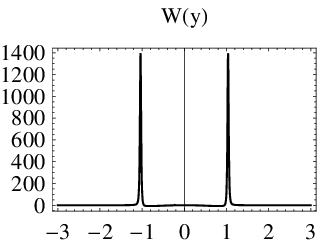}}
\subfigure[$f_R|_{y=0}=0.106$]{\label{PM55}
\includegraphics[width=11cm,height=5.5cm]{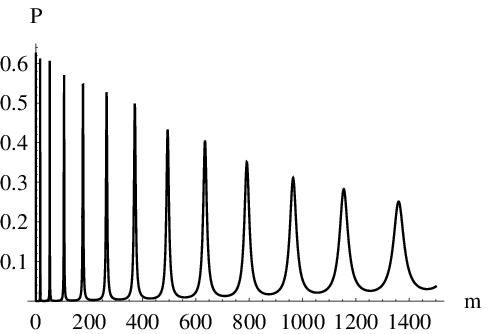}}
\end{center}
\caption{Plots of the function $f_R(y)$, effective potential $W(z(y))$, and relative probability $P(m^2)$ of the even resonance KK modes on the brane generated by K-field with $b=2,d=-2,\lambda=-0.6$, and $f_R|_{y=0}=0.106$.}
\end{figure*}

For the set of parameters $b=2,~d=-2,~\lambda=-0.6~$, and $f_R|_{y=0}=0.106$, we can see that $f_R$ almost vanishes at $y=\pm1$ (see Fig.~\ref{FR55}). We display the effective potential $W(z(y))$ in Fig.~\ref{W55} and relative probability $P(m^2)$ in Fig.~\ref{PM55} to prepare for the next section.

\section{The impact of $f(R)$-brane}
\label{analyse}

Now, we analyze how $f(R)$-brane model impacts gravitational resonances. Focusing on Eq.~(\ref{Wy}), the different parts of {the} potential $W(z(y))$ between the GR-brane and $f(R)$-brane are denoted by {$\Delta{W}$}:
\begin{eqnarray}
\Delta{W}=\frac{3aa'}{2}\frac{f'_R}{f_R}-\frac{1}{4}\Big(\frac{af'_R}{f_R}\Big)^2+\frac{1}{2}\frac{aa'f'_R+a^2f''_R}{f_R}.
\label{DeltaW}
\end{eqnarray}

{First of all, we analyze the model with only one canonical scalar field.} In this case, we have for the GR-brane
\begin{eqnarray}
-\frac{3}{2} A''=\phi'^2\geq0,
\end{eqnarray}
which results in a lumplike solution of the wrap factor.

For $f(R)$-brane, we have Eq.~(\ref{Cphi1}). Fixing the solution of {the} warp factor, if we expect some huge differences of the potentials $W(z(y))$ between $f(R)$-brane and GR-brane, the value of $\frac{f'_R}{f_R}$ or $\frac{f''_R}{f_R}$ must be large enough at some {points} to form deep wells or high barriers. Considering that $f(R)$ should be continuous and satisfy $f_R>0$, the most promising approach may be that $f_R$ approaches zero at some locations.

{Because both $f(R)$ and $f_R$ are even functions of the extra dimension $y$ and, generally, $f(R)$ should be a constant at infinity, we need only consider the situation that $f_R$ approaches a very small value at one or two locations (more locations makes no difference).}

From Eq.~(\ref{DeltaW}) we know that $\Delta{W}$ is also an even function. We assume $\Delta{W}$ has two infinitely deep potential wells (the same analysis is appropriate for barriers) symmetrically situated at {$y={\pm}y_0$}. For simplify, we {consider the case that} these points are {minima} of $f_R(y)$. So, we have $f'_R=0$ and $f''_R\geq0$ at these points. We plug $f'_R=0$ into Eq.~(\ref{Cphi1}) {at $y=\pm y_0$}, then we have $-3 A'' f_R \geq f''_R$.

Obviously, the first two terms of $\Delta{W}$ in Eq.~(\ref{DeltaW}) vanish at {$y=\pm y_0$}. Since $-3 A'' f_R \geq f''_R$, the {ranges} of $\Delta{W}$ at $y={\pm} y_0$ {are} both $0<\Delta{W}=\frac12\frac{a^2 f''_R}{f_R}<\frac{-3a^2 A''}{2} {=a^2 \phi'^2}$, which is finite. So, it is hard for {the} potential $W(z(y))$ to have deep potential wells or barriers around $y=\pm y_0$ even if $f_R|_{y=\pm y_0}$ approach zero. The same analysis also {applies} to the situation that $W(z(y))$ is supposed to have one infinite potential well at $y=0$.

As for the Bloch brane, although the solutions of {the} warp factor and scalar fields are not {the same as} the previous model, the restrictions on the warp factor and $f(R)$ are similar (see Eq.~(\ref{a28})). So, we also have the same conclusion.

Finally, we discuss why {the third $f(R)$-brane model has a} strong influence on {the} potential $W(z(y))$. Taking $f(R)=R$ in Eq.~(\ref{fR3}), we find {that} although {$\lambda$ is a small parameter}, it can lead to a non-monotonic warp factor from $y=0$ to $y=+\infty$. Therefore, the shape of {the} potential $W(z(y))$ of the $f(R)$-brane becomes complicated. Furthermore, obviously when $\lambda<0$, it is possible that {the function $f_R(y)$ approaches zero at some points}. In Fig.~\ref{FR55}, we see that when $f_R(y)$ gets close to zero, there are two high barriers at the corresponding points (Fig.~\ref{W55}) but the GR-brane never has such a kind of feature.

\section{Conclusion and discussion}
\label{Conclusion}

In this work we {discussed} gravitational resonances in various $f(R)$-brane models. For the branes generated by one or two canonical scalar fields, we {found} that the $f(R)$-brane and GR-brane are similar from the view point of gravitational resonance. Then we {investigated} the brane generated by a single K-field. For a positive $\lambda$ (or a negative $\lambda$ with the warp factor varying monotonically), the structure of gravitational resonance of $f(R)$-brane does not deviate much from the GR-brane. This result is similar to the case with the standard kinetic term. {When the parameter $\lambda$ is negative and the wrap factor is non-monotonic, the structure of the gravitational resonance of the $f(R)$-brane can seriously deviate from the GR-brane. The degree of deviation relies mostly on the value of the parameter {$f_R|_{y=0}$}.}

Finally, we analyzed why the $f(R)$-brane and GR-brane have similar gravitational resonance structures when the background scalar field is canonical. Our conclusion is that the similarity is due to the limitations $f_R>0$ and Eq.~(\ref{Cphi1}), which constrain strictly the potential $W(z(y))$ of the $f(R)$-brane on the whole space. For the cases with more scalar fields (such as the Bloch brane), one may draw the same conclusion, as long as all kinetic terms of the scalar fields are canonical. {However, for the brane generated by a K-field, the function $f_R(y)$ could approach zero at some locations. Therefore, the effective potential $W(z(y))$ in the $f(R)$-brane model may have plentiful structures and result in a series of gravitational resonances, which is very different from the GR-brane model.}

\section*{Acknowledgements}

This work was supported by the National Natural Science Foundation of China (Grants Nos. 11205074, 11375075, and 11522541), and the Fundamental Research Funds for the Central Universities {(Grant No. lzujbky-2015-jl1). Y.Z. was also supported by the scholarship Granted by the Chinese Scholarship Council (CSC).}

\providecommand{\href}[2]{#2}\begingroup\raggedright\endgroup

%

\begin{thebibliography}{10}

\bibitem{Kaluza1921}
T.~Kaluza, {\it {On the Problem of Unity in Physics}},  {\em Sitzungsber.
  Preuss. Akad. Wiss. Berlin Math. Phys.} 966 (1921)

\bibitem{Klein1926}
O.~Klein, {\it {Quantum theory and five-dimensional theory of relativity}},  {\em
  Z. Phys.} {\bf 37} (1926) 895

\bibitem{Akama1982}
K.~Akama, {\it {An early proposal of 'Brane World'}},  {\em Lect. Notes Phys.}
  {\bf 176} (1982) 267. [\href{http://arxiv.org/abs/hep-th/0001113}{{\tt
  hep-th/0001113}}]

\bibitem{RubakovShaposhnikov1983}
V.~A. Rubakov and M.~E. Shaposhnikov, {\it {Do we live inside a domain wall?}},
  {\em Phys. Lett. B} {\bf 125} (1983) 136

\bibitem{RubakovShaposhnikov1983a}
V.~A. Rubakov and M.~E. Shaposhnikov, {\it {Extra space-time dimensions: towards
  a solution to the cosmological constant problem}},  {\em Phys. Lett. B} {\bf
  125} (1983) 139

\bibitem{Visser1985}
M.~Visser, {\it {An exotic class of Kaluza-Klein models}},  {\em Phys. Lett. B}
  {\bf 159} (1985) 22

\bibitem{Squires1986}
E.~J. Squires, {\it {Dimensional reduction caused by a cosmological constant}},
   {\em Phys. Lett. B} {\bf 167} (1986) 286

\bibitem{Antoniadis1990}
I.~Antoniadis, {\it A possible new dimension at a few Tev},  {\em Phys. Lett.
  B} {\bf 246} (1990) 377

\bibitem{Arkani-HamedDimopoulosDvali1998a}
N.~Arkani-Hamed, S.~Dimopoulos, and G.~Dvali, {\it {The Hierarchy problem and
  new dimensions at a millimeter}},  {\em Phys. Lett. B} {\bf 429} (1998)
  263. [\href{http://arxiv.org/abs/hep-ph/9803315}{{\tt hep-ph/9803315}}]

\bibitem{RandallSundrum1999}
L.~Randall and R.~Sundrum, {\it {A large mass hierarchy from a small extra
  dimension}},  {\em Phys. Rev. Lett.} {\bf 83} (1999) 3370.
  [\href{http://arxiv.org/abs/hep-ph/9905221}{{\tt hep-ph/9905221}}]

\bibitem{RandallSundrum1999a}
L.~Randall and R.~Sundrum, {\it {An alternative to compactification}},  {\em
  Phys. Rev. Lett.} {\bf 83} (1999) 4690.
  [\href{http://arxiv.org/abs/hep-th/9906064}{{\tt hep-th/9906064}}]

\bibitem{GregoryRubakovSibiryakov2000b}
R.~Gregory, V.~A. Rubakov, and S.~M. Sibiryakov, {\it {Opening up extra
  dimensions at ultra-large scales}},  {\em Phys. Rev. Lett.} {\bf 84} (2000)
  5928. [\href{http://arxiv.org/abs/hep-th/0002072}{{\tt
  hep-th/0002072}}]

\bibitem{CsakiErlichHollowood2000a}
C.~Cs\'aki, J.~Erlich, and T.~J. Hollowood, {\it Quasilocalization of gravity
  on a brane by resonant modes},  {\em Phys. Rev. Lett.} {\bf 84} (2000)
  5932. [\href{http://arxiv.org/abs/hep-th/0002161}{{\tt
  hep-th/0002161}}]

\bibitem{DvaliGabadadzePorrati2000a}
G.~Dvali, G.~Gabadadze, and M.~Porrati, {\it {Metastable gravitons and infinite
  volume extra dimensions}},  {\em Phys. Lett. B} {\bf 484} (2000) 112.
  [\href{http://arxiv.org/abs/hep-th/0002190}{{\tt hep-th/0002190}}]

\bibitem{Witten2000}
E.~Witten, {\it {The Cosmological constant from the viewpoint of string
  theory}}. [\href{http://arxiv.org/abs/hep-ph/0002297}{{\tt hep-ph/0002297}}]

\bibitem{CsakiErlichHollowood2000}
C.~Csaki, J.~Erlich, and T.~J. Hollowood, {\it {Graviton propagators, brane
  bending and bending of light in theories with quasilocalized gravity}},  {\em
  Phys. Lett. B} {\bf 481} (2000) 107.
  [\href{http://arxiv.org/abs/hep-th/0003020}{{\tt hep-th/0003020}}]

\bibitem{GregoryRubakovSibiryakov2000a}
R.~Gregory, V.~A. Rubakov, and S.~M. Sibiryakov, {\it {Gravity and antigravity
  in a brane world with metastable gravitons}},  {\em Phys. Lett. B} {\bf 489}
  (2000) 203. [\href{http://arxiv.org/abs/hep-th/0003045}{{\tt
  hep-th/0003045}}]

\bibitem{PiloRattazziZaffaroni2000}
L.~Pilo, R.~Rattazzi, and A.~Zaffaroni, {\it {The Fate of the radion in models
  with metastable graviton}},  {\em J. High Energy Phys.} {\bf 0007} (2000)
  056. [\href{http://arxiv.org/abs/hep-th/0004028}{{\tt hep-th/0004028}}]

\bibitem{CsakiErlichHollowoodTerning2001}
C.~Csaki, J.~Erlich, T.~J. Hollowood, and J.~Terning, {\it {Holographic RG and
  cosmology in theories with quasilocalized gravity}},  {\em Phys. Rev. D} {\bf
  63} (2001) 065019. [\href{http://arxiv.org/abs/hep-th/0003076}{{\tt
  hep-th/0003076}}]

\bibitem{DvaliGabadadzePorrati2000}
G.~R. Dvali, G.~Gabadadze, and M.~Porrati, {\it {4D gravity on a brane in 5D
  minkowski space}},  {\em Phys. Lett. B} {\bf 485} (2000) 208.
  [\href{http://arxiv.org/abs/hep-th/0005016}{{\tt hep-th/0005016}}]

\bibitem{DeffayetDvaliGabadadze2002}
C.~Deffayet, G.~Dvali, and G.~Gabadadze, {\it {Accelerated universe from
  gravity leaking to extra dimensions}},  {\em Phys. Rev. D} {\bf 65} (2002)
  044023. [\href{http://arxiv.org/abs/astro-ph/0105068}{{\tt
  astro-ph/0105068}}]

\bibitem{KolanovicPorratiRombouts2003}
M.~Kolanovic, M.~Porrati, and J.-W. Rombouts, {\it {Regularization of brane
  induced gravity}},  {\em Phys. Rev. D} {\bf 68} (2003) 064018.
  [\href{http://arxiv.org/abs/hep-th/0304148}{{\tt hep-th/0304148}}]

\bibitem{GabadadzeShifman2004}
G.~Gabadadze and M.~Shifman, {\it {Softly massive gravity}},  {\em Phys. Rev. D}
  {\bf 69} (2004) 124032. [\href{http://arxiv.org/abs/hep-th/0312289}{{\tt
  hep-th/0312289}}]

\bibitem{RingevalRombouts2005}
C.~Ringeval and J.-W. Rombouts, {\it {Metastable gravity on classical
  defects}},  {\em Phys. Rev. D} {\bf 71} (2005) 044001.
  [\href{http://arxiv.org/abs/hep-th/0411282}{{\tt hep-th/0411282}}]

\bibitem{ShaposhnikovTinyakovZuleta2004}
M.~Shaposhnikov, P.~Tinyakov, and K.~Zuleta, {\it {Quasilocalized gravity
  without asymptotic flatness}},  {\em Phys. Rev. D} {\bf 70} (2004) 104019.
  [\href{http://arxiv.org/abs/hep-th/0411031}{{\tt hep-th/0411031}}]

\bibitem{Seahra2005}
S.~S. Seahra, {\it {Ringing the Randall-Sundrum braneworld: metastable gravity
  wave bound states}},  {\em Phys. Rev. D} {\bf 72} (2005) 066002.
  [\href{http://arxiv.org/abs/hep-th/0501175}{{\tt hep-th/0501175}}]

\bibitem{Seahra2005a}
S.~S. Seahra, {\it {Metastable massive gravitons from an infinite extra
  dimension}},  {\em Int. J. Mod. Phys. D} {\bf 14} (2005) 2279.
  [\href{http://arxiv.org/abs/hep-th/0505196}{{\tt hep-th/0505196}}]

\bibitem{CveticRobnik2008}
M.~Cvetic and M.~Robnik, {\it {Gravity trapping on a finite thickness domain
  wall: an analytic study}},  {\em Phys. Rev. D} {\bf 77} (2008) 124003.
  [\href{http://arxiv.org/abs/0801.0801}{{\tt arXiv:0801.0801}}]

\bibitem{Nollert1999}
H.-P. Nollert, {\it {Quasinormal modes: the characteristic `sound' of black
  holes and neutron stars}},  {\em Class. Quant. Grav.} {\bf 16} (1999)
  159

\bibitem{DeWolfeFreedmanGubserKarch2000}
O.~DeWolfe, D.~Z. Freedman, S.~S. Gubser, and A.~Karch, {\it {Modeling the fifth
  dimension with scalars and gravity}},  {\em Phys. Rev. D} {\bf 62} (2000)
  046008. [\href{http://arxiv.org/abs/hep-th/9909134}{{\tt hep-th/9909134}}]

\bibitem{Gremm2000}
M.~Gremm, {\it {Four-dimensional gravity on a thick domain wall}},  {\em Phys.
  Lett. B} {\bf 478} (2000) 434.
  [\href{http://arxiv.org/abs/hep-th/9912060}{{\tt hep-th/9912060}}]

\bibitem{KehagiasTamvakis2001}
A.~Kehagias and K.~Tamvakis, {\it {Localized gravitons, gauge bosons and chiral
  fermions in smooth spaces generated by a bounce}},  {\em Phys. Lett. B} {\bf
  504} (2001) 38. [\href{http://arxiv.org/abs/hep-th/0010112}{{\tt
  hep-th/0010112}}]

\bibitem{CsakiErlichHollowoodShirman2000}
C.~Csaki, J.~Erlich, T.~J. Hollowood, and Y.~Shirman, {\it {Universal aspects of
  gravity localized on thick branes}},  {\em Nucl. Phys. B} {\bf 581} (2000)
  309. [\href{http://arxiv.org/abs/hep-th/0001033}{{\tt hep-th/0001033}}]

\bibitem{DzhunushalievFolomeevMinamitsuji2010}
V.~Dzhunushaliev, V.~Folomeev, and M.~Minamitsuji, {\it {Thick brane solutions}},
   {\em Rep. Prog. Phys.} {\bf 73} (2010) 066901.
  [\href{http://arxiv.org/abs/0904.1775}{{\tt arXiv:0904.1775}}]

\bibitem{CruzGomesAlmeida2011}
W.~Cruz, A.~Gomes, and C.~Almeida, {\it {Graviton resonances on deformed
  branes}},  {\em Europhys. Lett.} {\bf 96} (2011) 31001.
  [\href{http://arxiv.org/abs/1110.3104}{{\tt arXiv:1110.3104}}]

\bibitem{CruzMalufSousaAlmeida2016}
W.~T. Cruz, R.~V. Maluf, L.~J.~S. Sousa, and C.~A.~S. Almeida, {\it {Gravity
  localization in sine-Gordon braneworlds}},  {\em Ann. Phys.} {\bf 364}
  (2016) 2. [\href{http://arxiv.org/abs/1412.8492}{{\tt
  arXiv:1412.8492}}]

\bibitem{BazeiaGomes2004}
D.~Bazeia and A.~R. Gomes, {\it {Bloch brane}},  {\em J. High Energy Phys.} {\bf
  05} (2004) 012

\bibitem{CruzSousaMalufAlmeida2014}
W.~Cruz, L.~Sousa, R.~Maluf, and C.~Almeida, {\it {Graviton resonances on
  two-field thick branes}},  {\em Phys. Lett. B} {\bf 730} (2014) 314.
  [\href{http://arxiv.org/abs/1310.4085}{{\tt arXiv:1310.4085}}]

\bibitem{XieYangZhao2013}
Q.-Y. Xie, J.~Yang, and L.~Zhao, {\it {Resonance mass spectra of gravity and
  fermion on Bloch branes}},  {\em Phys. Rev. D} {\bf 88} (2013) 105014.
  [\href{http://arxiv.org/abs/1310.4585}{{\tt arXiv:1310.4585}}]

\bibitem{Armendariz-PiconDamourMukhanov1999}
C.~Armendariz-Picon, T.~Damour, and V.~F. Mukhanov, {\it {K-inflation}},  {\em
  Phys. Lett. B} {\bf 458} (1999) 209.
  [\href{http://arxiv.org/abs/hep-th/9904075}{{\tt hep-th/9904075}}]

\bibitem{GarrigaMukhanov1999}
J.~Garriga and V.~F. Mukhanov, {\it {Perturbations in $K$-inflation}},  {\em
  Phys. Lett. B} {\bf 458} (1999) 219.
  [\href{http://arxiv.org/abs/hep-th/9904176}{{\tt hep-th/9904176}}]

\bibitem{Armendariz-PiconMukhanovSteinhardt2001}
C.~Armendariz-Picon, V.~F. Mukhanov, and P.~J. Steinhardt, {\it {Essentials of
  k-essence}},  {\em Phys. Rev. D} {\bf 63} (2001) 103510.
  [\href{http://arxiv.org/abs/astro-ph/0006373}{{\tt astro-ph/0006373}}]

\bibitem{AdamGrandiSanchez-GuillenWereszczynski2008}
C.~Adam, N.~Grandi, J.~Sanchez-Guillen, and A.~Wereszczynski, {\it K-fields,
  compactons, and thick branes},  {\em J. Phys. A} {\bf 41} (2008) 212004.
  [\href{http://arxiv.org/abs/0711.3550}{{\tt arXiv:0711.3550}}]

\bibitem{BazeiaGomesLosanoMenezes2009}
D.~Bazeia, A.~R. Gomes, L.~Losano, and R.~Menezes, {\it {Braneworld models of
  scalar fields with generalized dynamics}},  {\em Phys. Lett. B} {\bf 671}
  (2009) 402. [\href{http://arxiv.org/abs/0808.1815}{{\tt arXiv:0808.1815}}]

\bibitem{LiuZhongYang2010}
Y.-X. Liu, Y.~Zhong, and K.~Yang, {\it {Scalar-kinetic branes}},  {\em Europhys.
  Lett.} {\bf 90} (2010) 51001. [\href{http://arxiv.org/abs/0907.1952}{{\tt
  arXiv:0907.1952}}]

\bibitem{BazeiaLobLosanoMenezes2013}
D.~Bazeia, A.~S. Lob\~ao, L.~Losano, and R.~Menezes, {\it {First-order
  formalism for flat branes in generalized N-field models}},  {\em Phys. Rev.
  D} {\bf 88} (2013) 045001. [\href{http://arxiv.org/abs/1306.2618}{{\tt arXiv:1306.2618}}]

\bibitem{ZhongLiuZhao2014a}
Y.~Zhong, Y.-X. Liu, and Z.-H. Zhao, {\it {Nonperturbative procedure for stable
  $K$-brane}},  {\em Phys. Rev. D} {\bf 89} (2014) 104034.
  [\href{http://arxiv.org/abs/1401.0004}{{\tt arXiv:1401.0004}}]

\bibitem{BazeiaLobaoMenezes2015}
D.~Bazeia, A.~Lob\~ao, and R.~Menezes, {\it {Thick brane models in generalized
  theories of gravity}},  {\em Phys. Lett. B} {\bf 743} (2015) 98.
  [\href{http://arxiv.org/abs/1502.04757}{{\tt arXiv:1502.04757}}]

\bibitem{ZhongLiuZhao2014}
Y.~Zhong, Y.-X. Liu, and Z.-H. Zhao, {\it {Brane structure and metastable
  graviton in five-dimensional model with (non)canonical scalar field}}.
  [\href{http://arxiv.org/abs/1404.2666}{{\tt arXiv:1404.2666}}]

\bibitem{GuoLiuZhaoChen2012}
H.~Guo, Y.-X. Liu, Z.-H. Zhao, and F.-W. Chen, {\it {Thick branes with a
  non-minimally coupled bulk-scalar field}},  {\em Phys. Rev. D} {\bf 85}
  (2012) 124033. [\href{http://arxiv.org/abs/1106.5216}{{\tt
  arXiv:1106.5216}}]

\bibitem{Buchdahl1970}
H.~A. Buchdahl, {\it {Non-linear lagrangians and cosmological theory}},  {\em
  Mon. Not. R. Astron. Soc.} {\bf 150} (1970) 1

\bibitem{Starobinsky1980}
A.~A. Starobinsky, {\it {A new type of isotropic cosmological models without
  singularity}},  {\em Phys. Lett. B} {\bf 91} (1980) 99

\bibitem{Capozziello2002}
S.~Capozziello, {\it {Curvature  quintessence}},  {\em Int. J. Mod. Phys. D} {\bf
  11} (2002) 483. [\href{http://arxiv.org/abs/gr-qc/0201033}{{\tt
  gr-qc/0201033}}]

\bibitem{NojiriOdintsov2003d}
S.~Nojiri and S.~D. Odintsov, {\it {Modified gravity with negative and positive
  powers of the curvature: unification of the inflation and of the cosmic
  acceleration}},  {\em Phys. Rev. D} {\bf 68} (2003) 123512.
  [\href{http://arxiv.org/abs/hep-th/0307288}{{\tt hep-th/0307288}}]

\bibitem{CarrollDuvvuriTroddenTurner2004}
S.~M. Carroll, V.~Duvvuri, M.~Trodden, and M.~S. Turner, {\it {Is cosmic
  speed-up due to new gravitational physics?}},  {\em Phys. Rev. D} {\bf 70}
  (2004) 043528. [\href{http://arxiv.org/abs/astro-ph/0306438}{{\tt
  astro-ph/0306438}}]

\bibitem{NojiriOdintsov2006a}
S.~Nojiri and S.~D. Odintsov, {\it {Modified f(r) gravity consistent with
  realistic cosmology: from matter dominated epoch to dark energy universe}},
  {\em Phys. Rev. D} {\bf 74} (2006) 086005.
  [\href{http://arxiv.org/abs/hep-th/0608008}{{\tt hep-th/0608008}}]

\bibitem{AmendolaGannoujiPolarskiTsujikawa2007}
L.~Amendola, R.~Gannouji, D.~Polarski, and S.~Tsujikawa, {\it {Conditions for
  the cosmological viability of f(R) dark energy models}},  {\em Phys. Rev. D}
  {\bf 75} (2007) 083504. [\href{http://arxiv.org/abs/gr-qc/0612180}{{\tt
  gr-qc/0612180}}]

\bibitem{HuSawicki2007}
W.~Hu and I.~Sawicki, {\it {Models of f(R) cosmic acceleration that evade
  solar-system tests}},  {\em Phys. Rev. D} {\bf 76} (2007) 064004.
  [\href{http://arxiv.org/abs/0705.1158}{{\tt arXiv:0705.1158}}]

\bibitem{Starobinsky2007}
A.~A. Starobinsky, {\it {Disappearing cosmological constant in f(R) gravity}},
  {\em JETP Lett.} {\bf 86} (2007) 157.
  [\href{http://arxiv.org/abs/0706.2041}{{\tt arXiv:0706.2041}}]

\bibitem{DeTsujikawa2010}
A.~De~Felice and S.~Tsujikawa, {\it {f(R) theories}},  {\em Living Rev. Rel.}
  {\bf 13} (2010) 3. [\href{http://arxiv.org/abs/1002.4928}{{\tt
  arXiv:1002.4928}}]

\bibitem{SotiriouFaraoni2010}
T.~P. Sotiriou and V.~Faraoni, {\it {$f(R)$ theories of gravity}},  {\em Rev.
  Mod. Phys.} {\bf 82} (2010) 451. [\href{http://arxiv.org/abs/0805.1726}{{\tt
  arXiv:0805.1726}}]

\bibitem{NojiriOdintsov2011}
S.~Nojiri and S.~D. Odintsov, {\it {Unified cosmic history in modified gravity:
  from F(R) theory to Lorentz non-invariant models}},  {\em Phys. Rept.} {\bf
  505} (2011) 59. [\href{http://arxiv.org/abs/1011.0544}{{\tt
  arXiv:1011.0544}}]

\bibitem{NojiriOdintsov2000b}
S.~Nojiri and S.~D. Odintsov, {\it {Brane-world cosmology in higher derivative
  gravity or warped compactification in the next-to-leading order of ads/cft
  correspondence}},  {\em J. High Energy Phys.} {\bf 07} (2000) 049.
  [\href{http://arxiv.org/abs/hep-th/0006232}{{\tt hep-th/0006232}}]

\bibitem{NojiriOdintsovOgushi2002a}
S.~Nojiri, S.~D. Odintsov, and S.~Ogushi, {\it {Cosmological and black hole
  brane world universes in higher derivative gravity}},  {\em Phys. Rev. D} {\bf
  65} (2002) 023521. [\href{http://arxiv.org/abs/hep-th/0108172}{{\tt
  hep-th/0108172}}]

\bibitem{ParryPichlerDeeg2005}
M.~Parry, S.~Pichler, and D.~Deeg, {\it {Higher-derivative gravity in brane
  world models}},  {\em JCAP} {\bf 0504} (2005) 014.
  [\href{http://arxiv.org/abs/hep-ph/0502048}{{\tt hep-ph/0502048}}]

\bibitem{AfonsoBazeiaMenezesPetrov2007}
V.~I. Afonso, D.~Bazeia, R.~Menezes, and A.~Y. Petrov, {\it {$f(R)$-brane}},
  {\em Phys. Lett. B} {\bf 658} (2007) 71.
  [\href{http://arxiv.org/abs/0710.3790}{{\tt arXiv:0710.3790}}]

\bibitem{DeruelleSasakiSendouda2008}
N.~Deruelle, M.~Sasaki, and Y.~Sendouda, {\it {Junction conditions in $f(R)$
  theories of gravity}},  {\em Prog. Theor. Phys.} {\bf 119} (2008) 237.
  [\href{http://arxiv.org/abs/0711.1150}{{\tt arXiv:0711.1150}}]

\bibitem{BalcerzakDabrowski2010}
A.~Balcerzak and M.~P. Dabrowski, {\it {Brane $f(R)$ gravity cosmologies}},  {\em
  Phys. Rev. D} {\bf 81} (2010) 123527.
  [\href{http://arxiv.org/abs/1004.0150}{{\tt arXiv:1004.0150}}]

\bibitem{Bouhmadi-LopezCapozzielloCardone2010}
M.~Bouhmadi-Lopez, S.~Capozziello, and V.~F. Cardone, {\it {Cosmography of
  $f(R)$ - brane cosmology}},  {\em Phys. Rev. D} {\bf 82} (2010) 103526.
  [\href{http://arxiv.org/abs/1010.1547}{{\tt arXiv:1010.1547}}]

\bibitem{DzhunushalievFolomeevKleihausKunz2010}
V.~Dzhunushaliev, V.~Folomeev, B.~Kleihaus, and J.~Kunz, {\it {Some thick brane
  solutions in $f(R)$-gravity}},  {\em J. High Energy Phys.} {\bf 04} (2010)
  130. [\href{http://arxiv.org/abs/0912.2812}{{\tt arXiv:0912.2812}}]

\bibitem{HoffDias2011}
J.~Hoff~da Silva and M.~Dias, {\it {Five dimensional f(R) braneworld models}},
  {\em Phys. Rev. D} {\bf 84} (2011) 066011.
  [\href{http://arxiv.org/abs/1107.2017}{{\tt arXiv:1107.2017}}]

\bibitem{LiuZhongZhaoLi2011}
Y.-X. Liu, Y.~Zhong, Z.-H. Zhao, and H.-T. Li, {\it {Domain wall brane in
  squared curvature gravity}},  {\em J. High Energy Phys.} {\bf 06} (2011) 135.
  [\href{http://arxiv.org/abs/1104.3188}{{\tt arXiv:1104.3188}}]

\bibitem{LiuLuWang2012}
H.~Liu, H.~Lu, and Z.-L. Wang, {\it {f(R) gravities, killing spinor equations,
  'bps' domain walls and cosmology}},  {\em J. High Energy Phys.} {\bf 1202}
  (2012) 083. [\href{http://arxiv.org/abs/1111.6602}{{\tt arXiv:1111.6602}}]

\bibitem{BazeiaMenezesPetrovSilva2013}
D.~Bazeia, R.~Menezes, A.~Y. Petrov, and A.~da~Silva, {\it {On the many-field
  $f(R)$ brane}},  {\em Phys. Lett. B} {\bf 726} (2013) 523.
  [\href{http://arxiv.org/abs/1306.1847}{{\tt arXiv:1306.1847}}]

\bibitem{BazeiaLobaoMenezesPetrovSilva2014}
D.~Bazeia, A.~S. Lob\~ao, R.~Menezes, A.~Y. Petrov, and A.~da~Silva, {\it
  Braneworld solutions for f(R) models with non-constant curvature},  {\em
  Phys. Lett. B} {\bf 729} (2014) 127.
  [\href{http://arxiv.org/abs/1311.6294}{{\tt arXiv:1311.6294}}]

\bibitem{BazeiaLobaoLosanoMenezesOlmo2015}
D.~Bazeia, A.~Lob\~ao, L.~Losano, R.~Menezes, and G.~J. Olmo, {\it {Braneworld
  solutions for modified theories of gravity with nonconstant curvature}},  {\em
  Phys. Rev. D} {\bf 91} (2015), 124006.
  [\href{http://arxiv.org/abs/1505.06315}{{\tt arXiv:1505.06315}}]

\bibitem{ZhongLiuYang2011}
Y.~Zhong, Y.-X. Liu, and K.~Yang, {\it {Tensor perturbations of f(R)-branes}},
  {\em Phys. Lett. B} {\bf 699} (2011) 398.
  [\href{http://arxiv.org/abs/1010.3478}{{\tt arXiv:1010.3478}}]
  
\bibitem{XuZhongYuLiu2015}
Z.-G. Xu, Y.~Zhong, H.~Yu, and Y.-X. Liu, {\it {The structure of $f(R)$-brane
  model}},  {\em Eur. Phys. J. C} {\bf 75} (2015) 368.
  [\href{http://arxiv.org/abs/1405.6277}{{\tt arXiv:1405.6277}}]

\bibitem{Giovannini2001a}
M.~Giovannini, {\it {Gauge invariant fluctuations of scalar branes}},  {\em
  Phys. Rev. D} {\bf 64} (2001) 064023.
  [\href{http://arxiv.org/abs/hep-th/0106041}{{\tt hep-th/0106041}}]

\bibitem{Giovannini2002}
M.~Giovannini, {\it {Localization of metric fluctuations on scalar branes}},
  {\em Phys. Rev. D} {\bf 65} (2002) 064008.
  [\href{http://arxiv.org/abs/hep-th/0106131}{{\tt hep-th/0106131}}]

\bibitem{KobayashiKoyamaSoda2002}
S.~Kobayashi, K.~Koyama, and J.~Soda, {\it {Thick brane worlds and their
  stability}},  {\em Phys. Rev. D} {\bf 65} (2002) 064014.
  [\href{http://arxiv.org/abs/hep-th/0107025}{{\tt hep-th/0107025}}]

\bibitem{Giovannini2003}
M.~Giovannini, {\it {Scalar normal modes of higher dimensional gravitating
  kinks}},  {\em Class. Quant. Gravity} {\bf 20} (2003) 1063.
  [\href{http://arxiv.org/abs/gr-qc/0207116}{{\tt gr-qc/0207116}}]

\bibitem{ZhongLiu2013}
Y.~Zhong and Y.-X. Liu, {\it {Linearization of thick K-branes}},  {\em Phys.
  Rev. D} {\bf 88} (2013) 024017.
  [\href{http://arxiv.org/abs/1212.1871}{{\tt arXiv:1212.1871}}]

\bibitem{ZhongLiu2015a}
Y.~Zhong, Y.-X. Liu, {\it {Pure geometric thick $f(R)$-branes: stability and
  localization of gravity}}. [\href{http://arxiv.org/abs/1507.00630}{{\tt
  arXiv:1507.00630}}]

\bibitem{LiuYangZhaoFuDuan2009}
Y.-X. Liu, J.~Yang, Z.-H. Zhao, C.-E. Fu, and Y.-S. Duan, {\it {Fermion
  localization and resonances on a de sitter thick brane}},  {\em Phys. Rev. D}
  {\bf 80} (2009) 065019. [\href{http://arxiv.org/abs/0904.1785}{{\tt
  arXiv:0904.1785}}]

\bibitem{AlmeidaFerreiraGomesCasana2009}
C.~A.~S. Almeida, M.~M. Ferreira, Jr., A.~R. Gomes, and R.~Casana, {\it {Fermion
  localization and resonances on two-field thick branes}},  {\em Phys. Rev. D}
  {\bf 79} (2009) 125022. [\href{http://arxiv.org/abs/0901.3543}{{\tt
  arXiv:0901.3543}}]

\bibitem{SouzaFariaHott2008}
A.~de~Souza~Dutra, J.~de~Faria, A.C.~Amaro, and M.~Hott, {\it {Degenerate and
  critical Bloch branes}},  {\em Phys. Rev. D} {\bf 78} (2008) 043526.
  [\href{http://arxiv.org/abs/0807.0586}{{\tt arXiv:0807.0586}}]

\bibitem{CruzMalufAlmeida2013}
W.~T. Cruz, R.~V. Maluf, and C.~A.~S. Almeida, {\it {Kalb-Ramond field
  localization on the Bloch brane}},  {\em Eur. Phys. J. C} {\bf 73} (2013)
  2523. [\href{http://arxiv.org/abs/1303.1096}{{\tt arXiv:1303.1096}}]

\bibitem{CruzLimaAlmeida2013}
W.~T. Cruz, A.~R.~P. Lima, and C.~A.~S. Almeida, {\it {Gauge field localization
  on the Bloch Brane}},  {\em Phys. Rev. D} {\bf 87} (2013) 045018.
  [\href{http://arxiv.org/abs/1211.7355}{{\tt arXiv:1211.7355}}]

\bibitem{SouzaBritoHoff2014}
A.~de~Souza~Dutra, G.~P. de~Brito, and J.~M. H.~da Silva, {\it {Asymmetrical
  bloch branes and the hierarchy problem}},  {\em Europhys. Lett.} {\bf 108}
  (2014) 11001. [\href{http://arxiv.org/abs/1312.0091}{{\tt arXiv:1312.0091}}]

\bibitem{ZhaoLiuZhong2014}
Z.-H. Zhao, Y.-X. Liu, and Y.~Zhong, {\it {U(1) gauge field localization on a
  Bloch brane with Chumbes-Holf da Silva-Hott mechanism}},  {\em Phys. Rev. D}
  {\bf 90} (2014) 045031. [\href{http://arxiv.org/abs/1402.6480}{{\tt
  arXiv:1402.6480}}]

\bibitem{XieGuoZhaoDuZhang2015}
Q.-Y. Xie, H.~Guo, Z.-H. Zhao, Y.-Z. Du, and Y.-P. Zhang, {\it {Mass spectrum
  of fermion on Bloch branes with new scalar-fermion coupling}}.
  [\href{http://arxiv.org/abs/1510.03345}{{\tt arXiv:1510.03345}}]

\end{thebibliography}
\end{document}